\documentclass[traditabstract]{aa}
\usepackage{graphicx}                    % For eps figures, newer & more powerful
\usepackage{bm}
\usepackage{amsmath}
\usepackage{txfonts}                    % Change the \texttt command to courier style
\usepackage{natbib}                     % For citations: redefine \cite commands
\bibpunct{(}{)}{;}{a}{}{,}
 \usepackage{threeparttable}
 \usepackage{amssymb}                    % useful mathematical symbols
\def \i{{\rm i}}
\def\tauc{\tau_{\rm corr}}
\def\beg{\begin{eqnarray}}
\def\ende{\end{eqnarray}}
\def\lsim{\lower.4ex\hbox{$\;\buildrel <\over{\scriptstyle\sim}\;$}}
\def\gsim{\lower.4ex\hbox{$\;\buildrel >\over{\scriptstyle\sim}\;$}}

\newcommand{\Pm}{\mbox{Pm}}

\renewcommand{\vec}[1]{\mbox{\boldmath $#1$}}

\def \Om  {{\it \Omega}}

\def \etaT{\eta_ {\rm T}}

\def\gsim{\lower.4ex\hbox{$\;\buildrel >\over{\scriptstyle\sim}\;$}} %$
\def\lsim{\lower.4ex\hbox{$\;\buildrel <\over{\scriptstyle\sim}\;$}} %$

\def\div{{\rm div}}

\renewcommand{\vec}[1]{\mbox{\boldmath $#1$}}

\def\aap{Astronomy \& Astrophysics}

\def\ara\&a{Ann. Rev. Astronomy Astrophysics}

\def\apjs{Astrophys. J. Suppl.}

\begin{document}

%\begin{article}

%\begin{opening}

%\title{Oscillating dynamo models  with alpha effect nonlocal in time }
%\title{Oscillating dynamo models with an $\alpha$ effect nonlocal in time}
\title{Oscillating dynamo models with a time-nonlocal $\alpha$ effect}
%\title{The solar rotation law  influenced by inclined  large-scale magnetic fields }
\titlerunning{Alpha effect nonlocal in time}%%%%%%%%%%%%%%%%%%%%%%%%%%%%%%%%%%%%%%%%%%%%%%%%%%%
%% Authors Names
%
%\author{G.~R\"udiger \and M.~K\"uker }
\author{G.~R\"udiger\inst{1,2} \and M.~Schultz\inst{1}\and A. Bonanno\inst{1,3}}

%%%%%%%%%%%%%%%%%%%%%%%%%%%%%%%%%%%%%%%%%%%%%%%%%%%

%%%%%%%%%%%%%%%%%%%%%%%%%%%%%%%%%%%%%%%%%%%%%%%%%%%
%% Affiliations 
%
\institute{Leibniz-Institut f\"ur Astrophysik Potsdam (AIP), An der Sternwarte 16, D-14482 Potsdam, Germany
	%   email: gruediger@aip.de
	\and
	University of Potsdam, Institute of Physics and Astronomy, D-14476 Potsdam, Germany
	\and
	INAF, Catania Astrophysical Observatory, Via Santa Sofia 78, 95123 Catania, Italy
}

  %    \institute{Leibniz-Institut f\"ur Astrophysik Potsdam (AIP), An der Sternwarte 16, D-14482 Potsdam, Germany,
        %             email: gruediger@aip.de}

\date{Received; accepted}
%%%%%%%%%%%%%%%%%%%%%%%%%%%%%%%%%%%%%%%%%%%%%%%%%%%
 
%\abstract{
%	The traditional $\alpha$ effect is extended by the first temporal derivative, with the result that in turbulent flows the magnetic field entering the induction process is effectively sampled in the past. As a consequence, dynamo excitation becomes easier and the cycle times are prolonged compared with solutions obtained from the local standard formulation. The shape of the magnetic cycles is also modified, with rise times becoming shorter than decline times so that the temporal profiles acquire a characteristic ``sawtooth'' appearance.
%	Moreover, in the nonlocal formulation the induced magnetic-field amplitudes acquire an additional dependence on the correlation time. For fixed dynamo number, the field amplitude reaches a maximum at a certain correlation time. For correlation times larger than this peak value, the model can reproduce Waldmeier-type trends known from solar sunspot observations. If the actual correlation time fluctuates around this value, the long-term magnetic cycle can occasionally enter weak-cycle episodes reminiscent of grand minima such as the Maunder minimum. For smaller dynamo numbers close to the critical one, however, rather long correlation times are required for this behaviour to occur.
%}

\abstract{
	The traditional $\alpha$ effect is extended by the first temporal derivative, so that in turbulent flows the magnetic field entering the induction process is effectively sampled in the past. As a consequence, dynamo excitation becomes easier and cycle times are prolonged relative to solutions obtained from the local standard formulation. The temporal shape of the magnetic cycles is also modified, with rise times becoming shorter than decline times, so that the profiles acquire a characteristic ``sawtooth'' appearance.
	In the nonlocal formulation, the induced magnetic-field amplitudes acquire an additional dependence on the correlation time. For fixed dynamo number, the field amplitude reaches a maximum at a certain correlation time. For correlation times larger than this peak value, the model can reproduce Waldmeier-type trends known from solar sunspot observations. If the actual correlation time fluctuates around this value, the long-term magnetic cycle can occasionally enter weak-cycle episodes reminiscent of grand minima such as the Maunder minimum. For smaller dynamo numbers close to the critical one, however, rather long correlation times are required for this behaviour to occur.
}


%-------------------------------------------------
\maketitle
%%%%%%%%%%%%%%%%%%%%%%%%%%%%%%%%%%%%%%%%%%%%%%%%%%%
%% Sections
%
%\section{Introduction} \label{Section1}
%The variability of the solar magnetic cycle remains one of the most challenging problems in stellar magnetohydrodynamics. Despite the success of mean-field dynamo theory in explaining the 11-year solar cycle \citep{1955ApJ...122..293P,SK66,RA03,stix2012sun, 2020LRSP...17....4C},
%(Parker 1955; Stix 2012; Charbonneau 2020), 
%several observational features, such as the asymmetric shape of sunspot cycles, the Waldmeier relations between rise time, amplitude, and rise rate, and the occurrence of Grand Minima, still lack a complete theoretical explanation. 
%Similar complex behavior has been observed in other late-type stars displaying cyclic magnetic activity 
%\citep{OKG09,RCG17,L16}, suggesting
%that these properties may be a generic consequence of a complex nonlinear mechanism which drives the internal dynamo. 

\section{Introduction} \label{Section1}
The variability of the solar magnetic cycle remains one of the most challenging problems in stellar magnetohydrodynamics. Despite the success of mean-field dynamo theory in accounting for the 11-year solar cycle \citep{1955ApJ...122..293P,SK66,RA03,stix2012sun,2020LRSP...17....4C}, several observed properties, such as the asymmetric shape of sunspot cycles, the Waldmeier relations between rise time, amplitude, and rise rate, and the occurrence of grand minima, still lack a complete theoretical explanation. Similar behaviour has also been observed in other late-type stars displaying cyclic magnetic activity \citep{OKG09,RCG17,L16}, suggesting that these properties may be a generic consequence of a nonlinear mechanism operating in stellar dynamos.

In fact, in standard mean-field models, the turbulent electromotive force (EMF) is assumed to respond instantaneously to the mean magnetic field, with the $\alpha$ effect and turbulent diffusivity parameterized as local functions of the field strength. However, the assumption of vanishing correlation time is seldom justified in stellar convection zones, where turbulent eddies have finite memory. Departures from instantaneous coupling lead naturally to a non-local-in-time EMF, in which the $\alpha$-effect depends on both the mean field and its time derivative. 

%We shall demonstrate that in this case, the EMF may be written as
We show that in this case the EMF can be written, to leading order, as
\beg
    \mathcal{E} \propto {\alpha}\,(B-\tau_{\mathrm{corr}}\dot{B}),
\ende
 representing the delayed response of turbulent helicity to variations in the magnetic field. 
%This extension breaks time-reversal symmetry and gives rise to asymmetric magnetic cycles, reduced excitation thresholds, 
%and amplitude--period anticorrelations—features reminiscent of the solar Waldmeier relations.
This extension breaks time-reversal symmetry and gives rise to asymmetric magnetic cycles, reduced excitation thresholds, 
and amplitude--period anticorrelations—features reminiscent of the solar Waldmeier relations.

In the analytical work developed here, we derive the expression for the EMF within the second-order correlation approximation (SOCA), explicitly including its dependence on the temporal derivative of the mean field. We then explore the dynamo equations that result from this time-nonlocal $\alpha$ effect and show that they yield oscillatory solutions with sharp rises, slower decays, and amplitude--period trends qualitatively consistent with key features of the solar cycle. The same mechanism may also operate in other late-type stars, and may help to explain the asymmetric magnetic activity cycles observed across a wide range of rotation rates.

A complementary, data-driven perspective on this problem was recently presented in 
\citet{2025SoPh..300..116B}, where the evolution of the monthly sunspot number was analyzed using the 
Sparse Identification of Nonlinear Dynamical Systems (SINDy) framework. 
That study revealed that the solar magnetic cycle can be accurately represented by a low-order nonlinear oscillator 
dominated by a cubic term of the form $B_{\phi}\dot{B}_{\phi}^{2}$, implying that the effective $\alpha$-effect depends explicitly 
on the field derivative. 
%This finding provides empirical support for the non-local-in-time formalism developed here, 
%establishing a clear link between the observed solar-cycle dynamics and a physically motivated modification of mean-field theory.
This finding provides empirical support for the non-local-in-time formalism developed here, and suggests a possible link between the observed solar-cycle dynamics and a physically motivated modification of mean-field theory.

%Basing on observational results by \cite{RCG17,G19}       and \cite{WHL20} we shall in particular discuss the shape of the activity cycles with respect to their skewness which may  be desribed  as  the ratio
%\begin{eqnarray}
%\varepsilon= \frac{\tau_{\rm rise}}{\tau_{\rm dec}},
%\label{asymmetry}
%\end{eqnarray}
 %where ${\tau_{\rm rise}}$ is the time from the cycle  beginning  to the maximum and ${\tau_{\rm dec}}$  the time from the maximum to the next zero. While $\varepsilon=1$ stans for a symmetric half-cycle, values $\varepsilon<1$ describe cycles with a faster  rise than decline   as it has been shown as typical for the thousends of Kepler stars analyzed in the paper by  Reinhold et al. Willamo et al. give averaged values of $\varepsilon =0.625$  for the Sun and slightly larger values (but always  $\varepsilon<1$) for their sample of Mt. Wilson stars. A more precise parameter to define a correlation of the Waldmeier-type has been used by  \cite{CS08} and \cite{G19}, who consider the correlation of a rise rate (the gradient from year to year)   of a cycle and the amplitude which is found to form a robust positive correlation (``Waldmeier2'') which we shall clearly reproduce by our (simplified) nonlinear models. Its formal explanation is that both magnetic field amplitude and the rise time of the cycle grow with growing correlation time but the magnetic field amplitude grows faster. A negative correlation of magnetic field amplitude and the rise time (``Waldmeier1'') does not appear.

 In the following sections we present the theoretical derivation of the time-dependent $\alpha$ effect (Sect.~\ref{EMF}), its implementation in simplified dynamo equations, and the resulting cyclic solutions together with their observational implications. We show that this framework can account for asymmetric rise and decay in simplified solar and stellar cycle models, and can reproduce Waldmeier-type trends once the correlation time of the convection is introduced as an additional parameter.
 
 Motivated by the observational results of \cite{RCG17,G19} and \cite{WHL20}, we focus in particular on the skewness of activity cycles, which may be described by the ratio
 \begin{eqnarray}
 	\varepsilon= \frac{\tau_{\rm rise}}{\tau_{\rm dec}},
 	\label{asymmetry}
 \end{eqnarray}
 where ${\tau_{\rm rise}}$ is the time from the beginning of a cycle to its maximum and ${\tau_{\rm dec}}$ is the time from the maximum to the next zero. While $\varepsilon=1$ stands for a symmetric half-cycle, values $\varepsilon<1$ describe cycles with faster rise than decline, as found to be typical of the thousands of \textit{Kepler} stars analysed by Reinhold et al. Willamo et al. give average values of $\varepsilon =0.625$ for the Sun and slightly larger values (but still $\varepsilon<1$) for their sample of Mt. Wilson stars. A more precise Waldmeier-type measure has been used by \cite{CS08} and \cite{G19}, who considered the correlation between the rise rate of a cycle and its amplitude and found a robust positive correlation (``Waldmeier2''). Our simplified nonlinear models can reproduce this trend. In the present formulation, this behaviour arises because both the magnetic-field amplitude and the rise time increase with increasing correlation time, but the amplitude increases faster. By contrast, a negative correlation between magnetic-field amplitude and rise time (``Waldmeier1'') does not appear in the present model.
 
 A further implication of the nonlocal-in-time $\alpha$ effect is that the dynamo saturation is no longer controlled by the field amplitude alone, but also by the rate of temporal change of the large-scale field over the correlation time $\tau_{\rm corr}$. This suggests possible observational diagnostics. In the solar case, the efficiency of large-scale field regeneration may then depend on cycle phase, with the strongest suppression expected during epochs of rapid field evolution. One may therefore expect systematic variations in cycle skewness, in the relation between rise rate and amplitude, and possibly in the lag between activity maximum and polar-field reversal. In a Babcock--Leighton interpretation, such a picture would also suggest that source proxies based on active-region tilt or dipole moment may correlate not only with activity level, but also with its temporal derivative. In stellar activity records, the same effect could appear statistically through correlations between cycle asymmetry, amplitude, and period.

%%%%%%%%%%%%%%%%%%%%%%%%%%%%%%%%%%%%%%%%%%%%%%%%%%%%%%%%%%%%%%%%%%%%%%%%%%%%%%%
\section{Electromotive force}\label{EMF}

%%%%%%%%%%%%%%%%%%%%%%%%%%%%%%%%%%%%%%%%%%%%%%%%%%%%%%%%%%%%%%%%%%%%%%%
%The homogeneous mean magnetic field may be introduced in the form
We consider a homogeneous mean magnetic field of the form
\beg
{\bar B_j(t)}= \dot B_{j} t.
\label{Bj}
\ende
Since the field vanishes at $t=0$, the following calculations apply primarily to fields close to reversal times in oscillatory dynamos. Below we also discuss the extension to finite fields at $t=0$.
%As the field vanishes at $t=0$ the results of the following calculations mainly concern to  fields close to the reversal times of oscillating dynamos. Below we shall also provide results for finite fields at $t=0$.
Then
\beg
\frac{\partial {u_i}}{\partial t}- \eta \, \Delta {u_i} + \frac{1}{\rho}\frac{\partial {p}}{\partial x_i} = \frac{1}{\mu_0\rho}(b_{i,l}-b_{l,i})t \dot B_l 
\label{partu}
\ende
with $\div \vec{u} =\div \vec{b}=0.$

To determine the temporal contribution to the mean electromotive force
$\langle \vec{u}\times \vec{b} \rangle$ at linear order, it is sufficient to solve the induction equation
%To find the temporal  field variations on the mean electromotive force
%$\langle \vec{u}\times \vec{b} \rangle$ at linear order it is enough to solve
%the induction equation
\beg
\frac{\partial {b_i}}{\partial t}- \eta \, \Delta {b_i}
= t \dot B_{j} u_{i,j} 
\label{partb}
\ende
The Navier--Stokes equation for the fluctuation $\vec{u}$ remains homogeneous as long as only terms linear in $\dot B_j$ are retained. One can then write
%The Navier-Stokes equation for the fluctuation
%$\vec{u}$ remains homogeneous if only expressions linear in $\dot B_{j}$
%re envisaged. One can thus work with 
\begin{eqnarray}
u_i (\vec{x},t)= \int\!\!\!\!\int (\hat u_i^{(0)} (\vec{k},\omega) + \hat u_i (\vec{k},\omega)+t {\tilde u_i} (\vec{k},\omega))\nonumber\\
 e^{{\rm i}({\bf k}{\bf x}-\omega t)} {\rm d}\vec{k} \, {\rm d}\omega, 
\nonumber\\
b_i (\vec{x},t) = \int\!\!\!\!\int (\hat b_i (\vec{k},\omega)
 + t\tilde b_{i} (\vec{k},\omega)) e^{{\rm i}({\bf k}{\bf x} -\omega t)}
 {\rm d} \vec{k} \, {\rm d}\omega.
\label{uibi}
\end{eqnarray}
This short series expansion excludes terms of higher order in the mean field, which would enter through contributions proportional to $t^2$, $t^3$, and higher powers.
%We note that this short series expansion excludes all terms higher than $\bar B$ as these terms are running with $t^2, t^3...$. 
The result is 
\beg
\hat u_i =-  \frac{1}{\mu_0\rho}\frac{{\rm i}\vec{k \dot B}}{{(-{\rm i} \omega
 + \nu k^2)}^2} \hat b_i + \hat u_i^{(0)},
   \tilde u_{i}= \frac{1}{\mu_0\rho}\frac{{\rm i}\vec{k \dot B}}{-{\rm i} \omega
 + \nu k^2} \hat b_i.
\label{ui}
\ende
and
\beg
\hat b_i =- \frac{{\rm i}\vec{k \dot B}}{{(-{\rm i} \omega + \eta k^2)}^2} \hat u_j,
 \quad\quad  \tilde b_{i}= \frac{{\rm i}\vec{k \dot B}}{-{\rm i} \omega
 + \eta k^2} \hat u_i.
\label{bi}
\ende
Eliminating the magnetic fluctuations one finds
\beg
\hat u_i =L \hat u_i^{(0)}
\label{L}
\ende
with
\beg
L= \{1+ \frac{(\vec{k\dot B})^2}{\mu_0\rho
(-{\rm i} \omega + \nu k^2)^2(-{\rm i} \omega + \eta k^2)^2}\}^{-1}
\label{L1}
\ende
This expression already indicates a suppressing effect associated with the time-dependent field through the term ${\dot B}^2$.
%We note the suppressing action of a time-dependent magnetic field with ${\dot B}^2$. 
For homogeneous turbulence one
immediately finds
\begin{eqnarray}
&&{\cal E}_i(\vec{x},t)=\epsilon_{ijk} \int\!\!\!\!\int \langle(\hat u_j (\vec{k},\omega)\hat b_k(\vec{-k },-\omega) +\nonumber\\ &&t (\hat u_j (\vec{k},\omega)\tilde b_k(\vec{-k },-\omega)+ \tilde u_j (\vec{k},\omega)\hat b_k(\vec{-k },-\omega))\rangle
 % e^{{\rm i}({\bf k}{\bf x} -\omega t)}
 {\rm d} \vec{k} \, {\rm d}\omega.
\label{Ei}
\end{eqnarray}
Hence
\begin{eqnarray}
{\cal E}_i(\vec{x},t)={\rm i }\epsilon_{ijk} \int\!\!\!\!\int 
(\frac{\vec{k\dot B}}{({\rm i} \omega + \nu k^2)^2 }- \frac{\vec{k\bar B}}{({\rm i} \omega + \nu k^2)})\hat Q_{jk}{\rm d} \vec{k} \, {\rm d}\omega
\label{Ei1}
\end{eqnarray}
with
\begin{eqnarray}
Q_{i j}(\vec{x},t)= \int\!\!\!\!\int \hat Q_{ij}(\vec{k},\omega) 
 e^{{\rm i}({\bf k}{\bf x}-\omega t)} {\rm d}\vec{k} \, {\rm d}\omega, 
\label{Q1}
\end{eqnarray}
with $\hat Q_{ij}(\vec{k},\omega) $  the spectral tensor of the correlation tensor $Q_{ij}$. It is simply
\begin{eqnarray}
\hat Q_{i j}(\vec{x},t)= L L^* \hat Q_{ij}^{(0)}
\label{Q2}
\end{eqnarray}
($L L^*$ is real) and 
\begin{eqnarray}
\hat Q_{i j}^{(0)}(\vec{x},t)= (k^2\delta_{ij}-k_ik_j)Q_1 -\i \epsilon_{ijk} k_k Q_2,
\label{Q3}
\end{eqnarray}
where $Q_2$ represents the helicity of the turbulent flow. One finds 
\begin{eqnarray}
{\cal E}_i(\vec{x},t)= \frac{2}{3} \int\!\!\!\!\int 
\frac{k^2 L L^* Q_2}{\omega^2+\eta^2 k^4}(\eta k^2 \bar B_i -\frac{\eta^2 k^4-\omega^2}{\omega^2+\eta^2 k^4} \dot B_i)
{\rm d} \vec{k} \, {\rm d}\omega
\label{Ei2}
\end{eqnarray}
For helical turbulence with $Q_2\neq 0$, an $\alpha$ effect therefore exists that is proportional to both $\bar{\vec B}$ and $\dot{\vec B}$. Hence, even at reversal, when $\bar B=0$, a residual contribution to the $\alpha$ effect remains.
%Obviously, if for helical turbulence with $Q_2\neq 0$  an alpha-effect  exists proportionate to $\bar {\vec B} $ and $\dot {\vec B} $,  so that even in case of a reversal, i.e. $\bar B=0$, a (small) alpha-effect remains. 
Without magnetic quenching ($L L^*=1$) it results
\begin{eqnarray}
{\cal E}_i(\vec{x},t)= \frac{2}{3} \int\!\!\!\!\int 
\frac{\eta k^2  Q_2 \bar B_i + \omega \dot B_i \partial Q_2/\partial \omega}{\omega^2+\eta^2 k^4}
{\rm d} \vec{k} \, {\rm d}\omega.
\label{Ei2a}
\end{eqnarray} 
The new term involving $\dot B$ is negative for monotonically decreasing $Q_2$ and vanishes for white-noise turbulence with infinitely short correlation time.
% The new term with $\dot B$ is obviously negative for monotonously decreasing $Q_2$ and vanishes  for white noise turbulence with infinitely short correlation time. 
The latter statement does not hold for the first term in (\ref{Ei2}) representing the classical $\alpha$ effect. If, on the other hand, the spectrum of $Q_2$ is so steep that the approximation $Q_2\simeq \delta(\omega)$ holds then the factor in front of $\bar {\vec B}$ is $\eta k^2 \simeq 1/\tau_{\rm corr}$ as often used in large-eddy simulations. 
%It makes thus sense to write
It is therefore natural to write
 \begin{eqnarray}
{\vec{\cal E}}(\vec{x},t)\propto\alpha (\bar {\vec B}(t) - \tau_{\rm corr} \dot {{\vec B}}).
\label{Ei3}
\end{eqnarray}
 
 The magnetic suppression of the $\alpha$ effects  here runs with $L L^*$, i.e.
\begin{eqnarray}
{\vec{\cal E}}(\vec{x},t)=\frac{\alpha}{1 + \gamma {\dot B}^2} (\bar {\vec B}(t) - \tau_{\rm corr} \dot {{\vec B}}).
\label{Ei4}
\end{eqnarray}
By the same argument as above, the coefficient $\gamma$ is expected to be of order $\tau_{\rm corr}^2$.
%By the same argumentation as above the coefficient $\gamma$ proves to be of order $\tau_{\rm corr}^2$.

%The standard quenching of the EMF with the magnetic energy ${\bar B}^2$ does not appear in this expression as we developed  (\ref{uibi}) only until the first order of $t$. If the series included $t^2$ and $t^3$ then the quenching terms also with ${\bar B}^2$ will appear.
%%%%%%%%%%%%%%%%%%%%%%%%%%%%%%%%%%%%%%
The standard quenching of the EMF with the magnetic energy ${\bar B}^2$ does not appear in this expression because the expansion in Eq.~(\ref{uibi}) has been carried out only to first order in $t$. If the series were extended to include terms of order $t^2$ and $t^3$, then quenching terms involving ${\bar B}^2$ would also appear.

\section{Dynamo waves}
%%%%%%%%%%%%%%%%%%%%%%%%%%%%%%%%
%The simplest dynamo wave equations are have been given by Parker (1975). 
The simplest dynamo-wave equations were given by Parker (1975).
In plane Cartesian geometry, we consider a mean magnetic field subject to a shear flow and an $\alpha$ effect. All quantities vary only in the $z$ direction, with fixed wave number $K=1$. Time is normalized with the magnetic-diffusion time, and the magnetic field is normalized with its equipartition value $B_{\rm eq}=\mu_0 \rho \langle u^2\rangle$. The resulting time-dependent amplitude equations are
%In plane Cartesian geometry there is a mean magnetic field subject to a (strong) shear
%flow and an $\alpha$ effect. All quantities vary only in the $z$-direction with the fixed wave number $K=1$. The time coordinate is normalized with the magnetic-diffusion time and the magnetic fields are normalized with its equipartition value $B_{\rm eq}=\mu_0\rho \langle u^2\rangle$. Then the time-dependent amplitude equations are
\begin{eqnarray}
\begin{aligned}
\dot A + A &=  C_\alpha  ( B - \tauc \dot B), \\ 
\dot B + B&=  \i C_\Om A + C_\alpha(A - \tauc \dot A)
\end{aligned}
\label{D1}
\end{eqnarray}
The real part of $B$ is interpreted as the azimuthal magnetic field $B_y$, while the real part of $A$ is interpreted as the latitudinal field $B_x$. Because Eq.~(\ref{D1}) depends linearly on $\dot B$, one generally expects growing and decaying phases of an oscillatory cycle to become asymmetric.
%The real part of $B$ is considered as the invisible azimuthal magnetic field $B_y$ and the real part of $A$ is considered as the latitudinal field $B_x$. Because of the linearity of (\ref{D1}) in $\dot B$ any cycle in oscillating dynamos will loose its symmetry for growing and decaying fields. 
%%%%%%%%%%%%%%%%%%%%%%%%%%%%%%%%%%%%%%%%
\subsection{A linear model}

We start with the linear problem. Equation~(\ref{D1}) then admits two eigenfrequencies, one of which corresponds to a decaying mode.
%We start with the linear problem. Then Eq. (\ref{D1}) possesses { two} eigenfrequencies as solution from which one is a decaying mode.  
For Fourier modes $e^{ \omega t}$ one finds
\begin{eqnarray}
\omega^2 + \omega \frac{2+2\tauc C_\alpha^2+ \i \tauc C_\alpha C_\Om}{1-\tauc^2 C_\alpha^2}  + 
\frac{1-C_\alpha^2 - \i C_\alpha C_\Om}{1-\tauc^2 C_\alpha^2}=0.
\label{D4}
\end{eqnarray}
The solutions of this equation may be written as $\omega= \omega_{\rm gr}  +  \i \omega_{\rm osc}$. Then we find  neutral solutions for $\omega_{\rm gr}= 0$. 

The condition $C_\alpha=1/\tauc$ plays an important role in what follows, as it separates two qualitatively different solution regimes of Eqs.~(\ref{D1}).
%We note that the condition $C_\alpha=1/\tauc$ plays an importanf role in our theory, we shall demonstrate that it separates the range of stable and unstable solutions of the Eqs. (\ref{D1}). 
For $\tauc C_\alpha>1$, the terms on the right-hand side proportional to $C_\alpha$ dominate the character of the system. In this regime, the relative minus sign between $A$ and $\dot A$, and between $B$ and $\dot B$, acts formally like a negative diffusivity, so that solutions with $C_\alpha>1/\tauc$ become unstable in this simplified model.
%For $\tauc C_\alpha>1$ the character of the equation system (\ref{D1}) is determined by its RHS terms running with $C_\alpha$. The minus between $A$ and $\dot A$ and between $B$ and $\dot B$ plays the role of a negative diffusivity hence all  solutions  with $C_\alpha>1/\tauc $ are unstable. 

The solution of (\ref{D4}) is
\begin{equation}
\begin{split}
&\omega=   - \frac{1+\tauc C_\alpha^2 +\i \tauc C_\alpha C_\Om/2 }{1-\tauc^2 C_\alpha^2}\pm\\\
  &\frac{\sqrt{C_\alpha^2 (1+2 \tauc)+\i C_\alpha C_\Om(1+\tauc)+\tauc^2 C_\alpha^2(1- C_\alpha^2)}}{(1-\tauc^2 C_\alpha^2)} .
 \end{split}
\label{D6}
\end{equation}
Hence
\begin{eqnarray}
\begin{split}
\omega= &  -  (1+\tauc C_\alpha^2+ \frac{\i}{2}\tauc C_\alpha C_\Om) \\\ &\pm \sqrt{C_\alpha^2(1+2\tauc)+\i C_\alpha C_\Om(1+\tauc)},
 \end{split}
\label{D7}
\end{eqnarray}
if higher-order terms from $\tauc^2$ on are neglected. Consequently,
\begin{eqnarray}
\omega_{\rm gr}= (C_\alpha -1)(1-\tauc C_\alpha)+...
\label{D8}
\end{eqnarray}
and 
\begin{eqnarray}
\omega_{\rm osc}= \frac{C_\Om}{2}(1-\tauc C_\alpha)+...
\label{D9}
\end{eqnarray}
for $\tauc C_\alpha<1$. The second frequency belongs to a  decaying  mode for small $C_\alpha$ with $\omega_{\rm gr}<0$.

According to Eq.~(\ref{D8}), the growth rate vanishes at $C_\alpha=1$ and becomes formally negative once $\tauc C_\alpha>1$.
%After (\ref{D8}) the growth rate vanishes for $C_\alpha=1$ and it becomes formally negative for $\tauc C_\alpha>1$.  
The true growth rate for $C_\alpha=1/\tauc$ results from (\ref{D6}) as $\omega_{\rm gr}= (1-\tauc)/2\tauc$ as the only value.

\begin{figure}
\centerline{
\vbox{
\includegraphics[width=1.0\columnwidth]{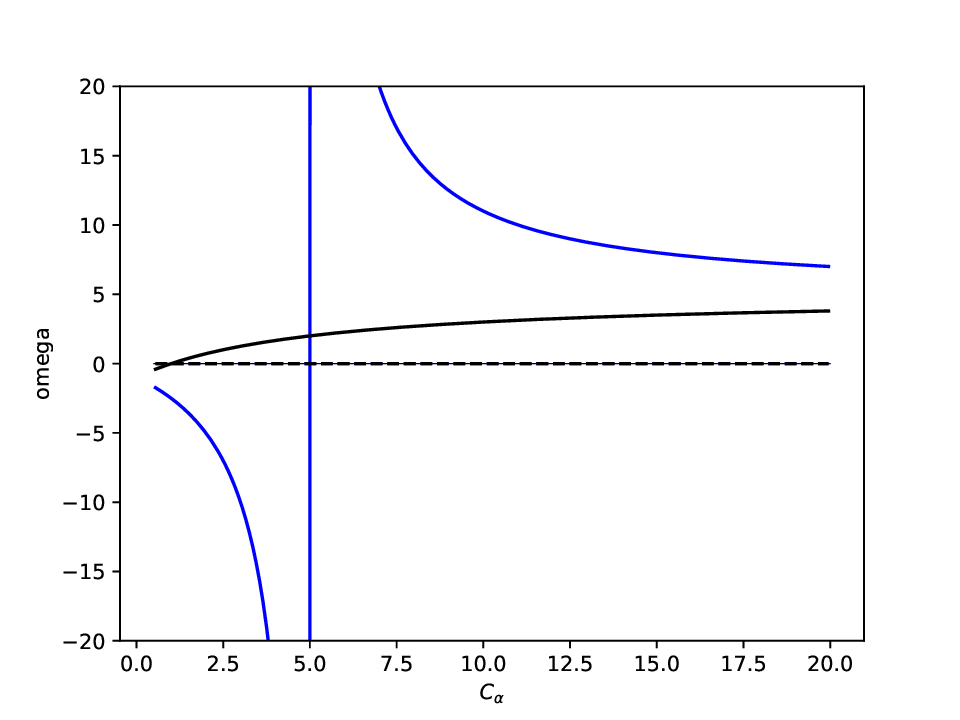}
\includegraphics[width=1.0\columnwidth]{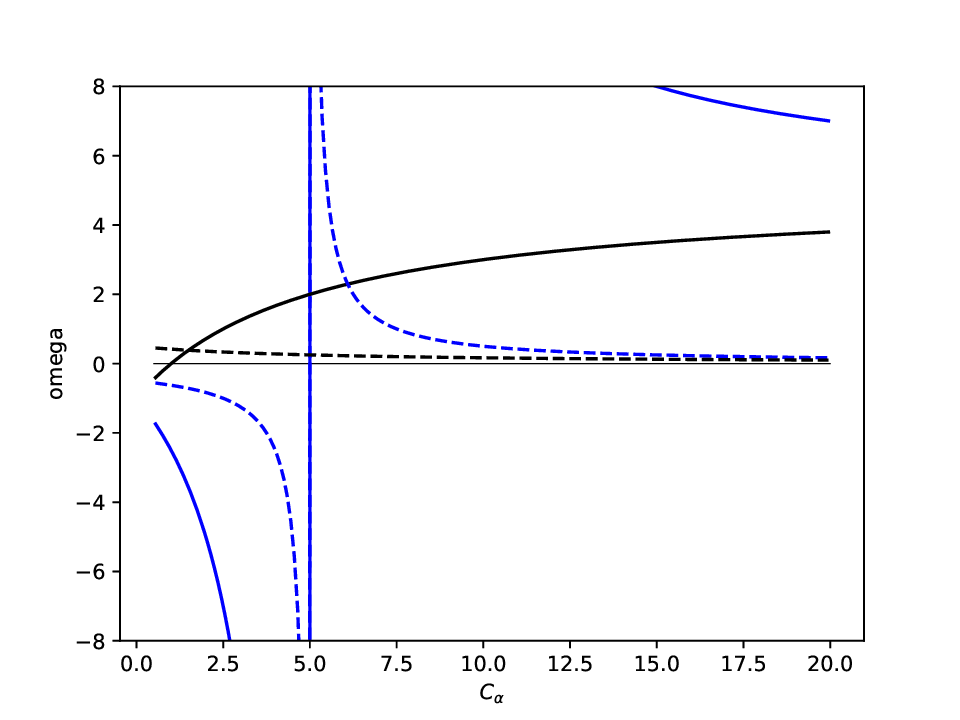}
}}
\caption{The real parts (solid lines) and the imaginary parts (dashed lines, 10x magnification) of the  two solutions of the dispersion relation  (\ref{D6}) for the linear model (\ref{D1}) for $C_\Om=0$ (top) and $C_\Om=0.1$ (bottom). Always $\tauc=0.2$.
}.
\label{fig0}
\end{figure}

In the framework of Eq.~(\ref{D1}), the nonlocal $\alpha$ effect therefore permits self-excited dynamo solutions only in the interval $1 \leq C_\alpha \leq 1/\tauc$, in contrast to the  local formulation with $\tauc=0$. For $C_\alpha>1/\tauc$, the corresponding solutions are unstable in the present model and can only be reached if the initial conditions are chosen very close to the final state.
%The nonlocal $\alpha$ effect expression in Eq. (\ref{D1}), therefore,  allows stable  dynamo solutions  only  for $ 1\leq C_\alpha \leq 1/\tauc$ in contrast to the standard local formulatiomn with $\tauc=0$. The solutions for $C_\alpha>1/\tauc$ prove to be  unstable, they only appear if the initial conditions are very close to the final solution. 
We note, that $C_\Om$ does not appear in Eq. (\ref{D8}) so that it does not influence the $ C_\alpha$ for neutral excitation.  We have numerically  solved Eq. (\ref{D4}). Fig. \ref{fig0} gives the result for $C_\Om=0$ and $C_\Om=0.1$. The black line gives the growing standard mode for $C_\alpha$ exceeding the critical value while the blue line gives  a decaying mode for $C_\alpha<1/\tauc$ and an extra growing mode for $C_\alpha >1/\tauc$. This extra mode is steady for $C_\Om=0$ and it is oscillating for $C_\Om=0.1$. 

In the latter case (with differential rotation), for $\tauc>1/C_\alpha$ two oscillatory branches appear: a slowly oscillating solution with low growth rate and a more rapidly oscillating solution with higher growth rate, reached for different initial conditions. In the presence of fluctuations of $\tauc$, or of random source terms in the induction equations (\ref{D1}), transitions between these branches may in principle occur.
%In the latter case (with differential rotation) for  $\tauc> 1/C_\alpha$ there are two oscillating solutions: a slowly oscillating solution with low growth rate and  a fast oscillating solution with a higher growth rate which are realized for different initial conditions. Random transitions between the two stable solutions should be possible under the influence of fluctuating $\tauc$ and or the influence  of random source terms in the induction euations (\ref{D1}). 
The plot for $C_\Om=0$ in Fig. \ref{fig0} demonstrates that such transition to an oscillating solution is not possible for $C_\Om=0$ in contradiction to the slightly different models discussed by \cite{Meinel90} and \cite {Hoyng90}. In our model fast magnetic reversals are also possible but only if a weak differential rotation exists and if the product $\tauc C_\alpha$ exceeds unity. 
%We note that the model does not provide oscillatory solutions with the local standard formulation with $\tauc=0$.
We again underline that, within this simplified formulation, the local standard case with $\tauc=0$ does not provide oscillatory solutions.

The solutions of (\ref{D1}) oscillate even for the smallest values of $C_\Om$. For small $C_\Om$ the periods of such dynamos are rather long and they become even longer for finite $\tauc$. The actual cycle frequency is not only determined by the diffusivity frequency but also on the value of the correlation time. For the rather large $\tauc=1$ the cycle time is twice the period for $\tauc=0$ which is $4\pi/C_\Om$.

That all $\alpha^2\Om$ dynamos oscillate and that for small $C_\Om$ the $C_\Om$ does not modify the critical $C_\alpha$ is also true for the more complex model of an disk dynamo with $\alpha$ effect asymmetric in $z$ \citep{RK13}..

%%%%%%%%%%%%%%%%%%%%%%%
%%%%%%%%%%%%%%%%%%%%%%%
\subsection{Nonlinear models}
%%%%%%%%%%%%%%%%%%%%%%%%%

The amplitudes of the oscillating field components are determined only in the nonlinear model.
%The amplitudes of the oscillating field components only result from nonlinear model calculations.
We begin with the nonlinear model for turbulence with a stationary correlation time $\tauc$.
%We start to discuss the nonlinear model for a turbulence with a stationary correlation time  $\tauc$. 
The actual time is normalized with  the diffusion frequency hence the time unit is the diffusion time, so that (say) $\tauc=1$ is a rather long correlation time. 
The cycle length of the model with $\tauc=0$ is the reference value. 
%The results for this  model and those with much longer correlation times are given in Fig. \ref{fig3a}. One finds for $\tauc=0.9$ an oscillation with a total cycle length of about 9 diffusion times which is a drastic amplification compared with the normalized  cycle length of $2\pi$ of the linear model with $\tauc=0$. 
The formulation (\ref{Bj}) is, however, incomplete.
%The formulation (\ref{Bj}) is certainly incomplete. 
Indeed,  the ansatz 
\beg
{\bar B_j(t)}= B_{0,j}+ \dot B_{j} t
\label{B0j}
\ende
%instead of (\ref{Bj}) also describes the field behaviour for finite magnetic field values.
instead of (\ref{Bj}) also allows for finite magnetic-field values.
 Equations (\ref{D11}) now read
\begin{eqnarray}
\dot A + A =C_\alpha  (\Phi B - \hat \Phi\tauc \dot B), \nonumber\\ \dot B + B= \i C_\Om A + C_\alpha(\Phi A - \hat \Phi\tauc \dot A)
\label{D11}
\end{eqnarray}
with the non-negative quenching functions
 \begin{eqnarray}
\Phi= 3 \int _0^1 \frac{((1+x^2{B}^2)^2 + \tauc^2 x^2{\dot B}^2) x^2}{ \left((1-x^2 B^2)^2 + x^2(2 B-\tauc \dot B)^2 \right)^2}  {\rm d} x
\label{D12}
\end{eqnarray}
and
\begin{eqnarray}
\hat \Phi =3 \int_0^1 \frac{(1+x^2 B^2)^2 x^2}{ \left((1-x^2 B^2)^2 +x^2 (2 B-\tauc \dot B)^2 \right)^2} {\rm d} x
\label{D13}
\end{eqnarray}
(see Appendix).
\begin{figure}
\centerline{
\vbox{
\includegraphics[width=1.0\columnwidth]{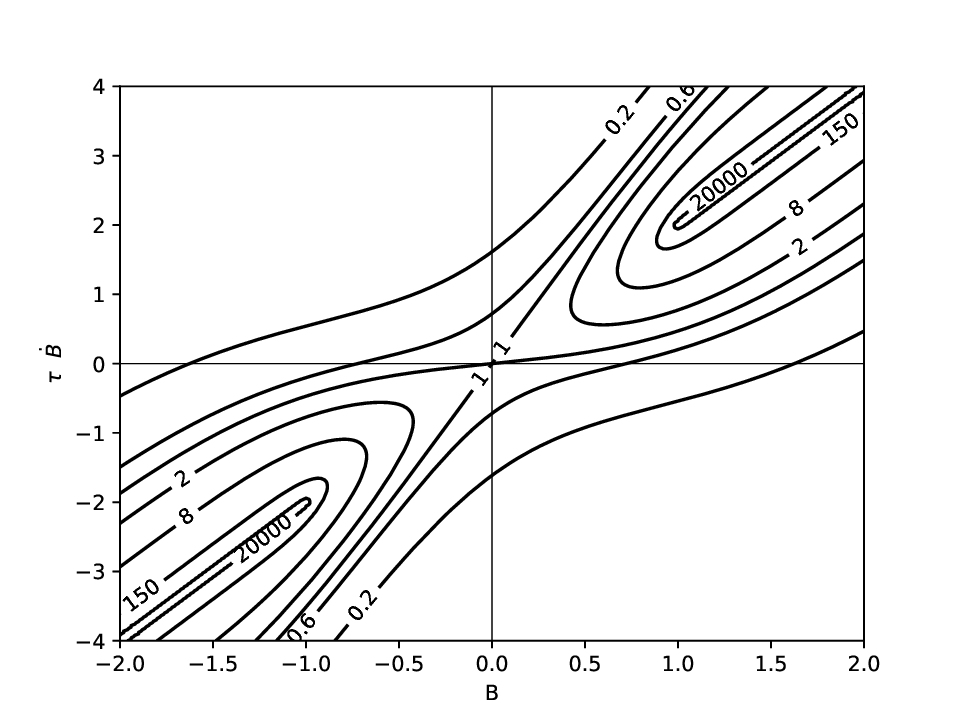}
\includegraphics[width=1.0\columnwidth]{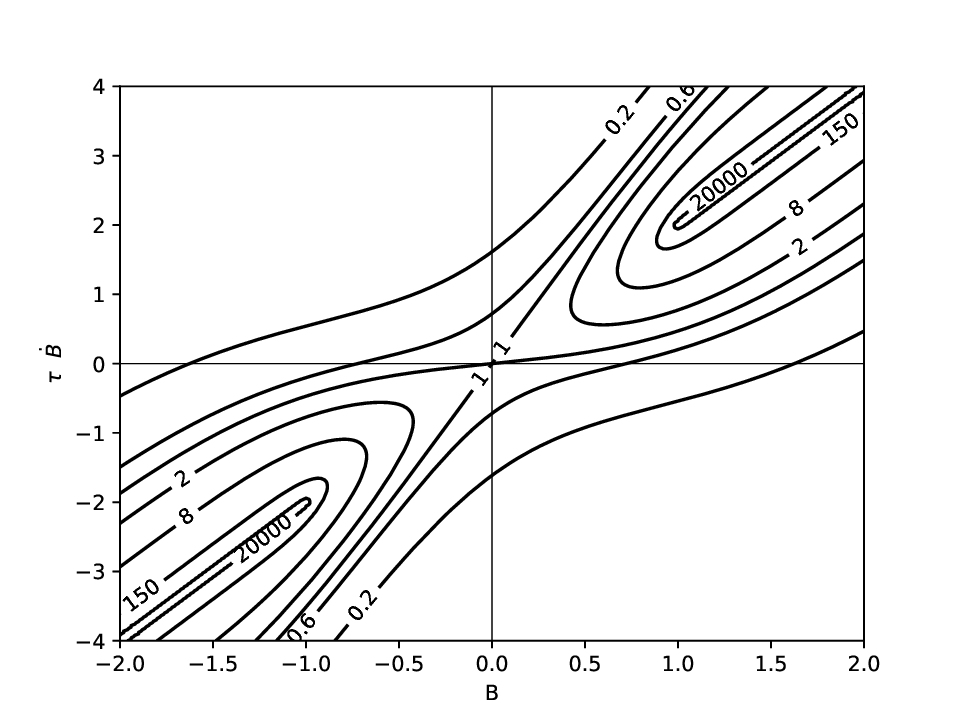}
}}
\caption{Isolines of $\Phi$ (top) and $\hat\Phi$ (bottom) in dependence of $B$ and $\tauc \dot B$.  For $\tauc=0$ (horizontal axis) one finds decreasing $\Phi$ (``quenching'') while  a maximum exists (``antiquenching'') for oscillating dynamos along the diagonal of this plot.
}.
\label{fig00}
\end{figure}
We note that the functions $\Phi$ and $\hat\Phi $ are both positive semi-definite, but they are not invariant under the transformation $\dot B \to -\dot B$. One therefore expects the two wings of a given half-cycle to become asymmetric. Since, however, the  functions remain invariant under the simultaneous transformation $\dot B \to -\dot B$ and $B \to -B$, a positive half-cycle has the same form as the following negative half-cycle. For fixed $\tauc$, the structure of the quenching functions does not in itself introduce cycle-to-cycle fluctuations of the amplitudes or cycle lengths.

%We note that both the functions $\Phi$ are positive definite but they are not invariant against the transformation $\dot B \to -\dot B$. In consequence, the two wings of one and the same half-cycle can no longer be symmetric. As, however, the quenching functions are invariant against te simultaneous transformation  $\dot B \to -\dot B$ and  $B \to - B$ a positive half-cycle has the same  form as the following negative half-cycle. The form of the quenching functions exclude the occurrence of regular or stochastic fluctuations of the field amplitudes and/or the cycle lengths.
Figure \ref{fig00} shows the isolines of  $\Phi$ (solid lines) and $\hat\Phi$ (dashed lines). Along the horizontal axis with $\tauc=0$ the $\Phi$ sinks with growing magnetic field as is known from the local  quenching theory, with $\Phi=O(B^{-3})$  \citep{M73,R74}. For finite correlation time, however, the $\Phi$ grows for positive $\dot B$ and sinks for negative $\dot B$. The $\alpha$ effect is thus quenched for negative $\dot B$ and it is antiquenched for postive $\dot B$. This is the mathematical background for the asymmetry of the rising phase and the falling phase of a half-cycle of oscillating dynamos.

The quantity $\tauc$ in (\ref{D11}) represents the correlation time normalized with the diffusion time.
%The quantity $\tauc$ in (\ref{D11}) represents the  correlation time normalized with the diffusion time. 
Hence, $\tauc = \etaT K^2 \tau = 4\pi^2 (u_{\rm rms}\tau/\ell)^2 (\ell/L)^2$, where $\ell$ and $\tau$ are the physical correlation length and time, respectively, and $L$ is a characteristic size of the dynamo domain.
%Hence $\tauc = \etaT K^2 \tau = 4\pi^2 (u_{\rm rms} \tau/\ell)^2 (\ell/L)^2$ where $\ell$ and $\tau$ are the physical correlation time and length, resp. and $L$ gives a characteristic size of the dynamo box. 
%With a Strouhal number of order unity we find $\tauc = O(\ell^2/L^2$ which for giant cells results of order unity or less.
For a Strouhal number of order unity, one finds $\tauc = O(\ell^2/L^2)$, which for giant cells is of order unity or smaller.

%%%%%%%%%%%%%%%%%%%%%%%%%%%%%%%%%%%%%%%%%

\section{Strong shear models}
%%%%%%%%%%%%%%%%%%%%%%%%%%%%%%%%
The time-dependent amplitude equations are
\begin{eqnarray}
\dot A + A =  C_\alpha \Phi (B - \tauc \dot B), \ \ \ \ \ \  \dot B + B= \i C_\Om A
\label{DD1}
\end{eqnarray}
We note that the neglect of the field component $A$ in this relation is only allowed for $\alpha\Omega$ dynamos with $C_\Omega\gg C_\alpha$. Hence,
\begin{eqnarray}
\ddot B + 2 \dot B + B = \i D \Phi (B- \tauc \dot B)
\label{DD3}
\end{eqnarray}
with the dynamo number $D= C_\alpha C_\Om$. 
\begin{figure*}
\centerline{
\hbox{
\includegraphics[width=1.1\columnwidth]{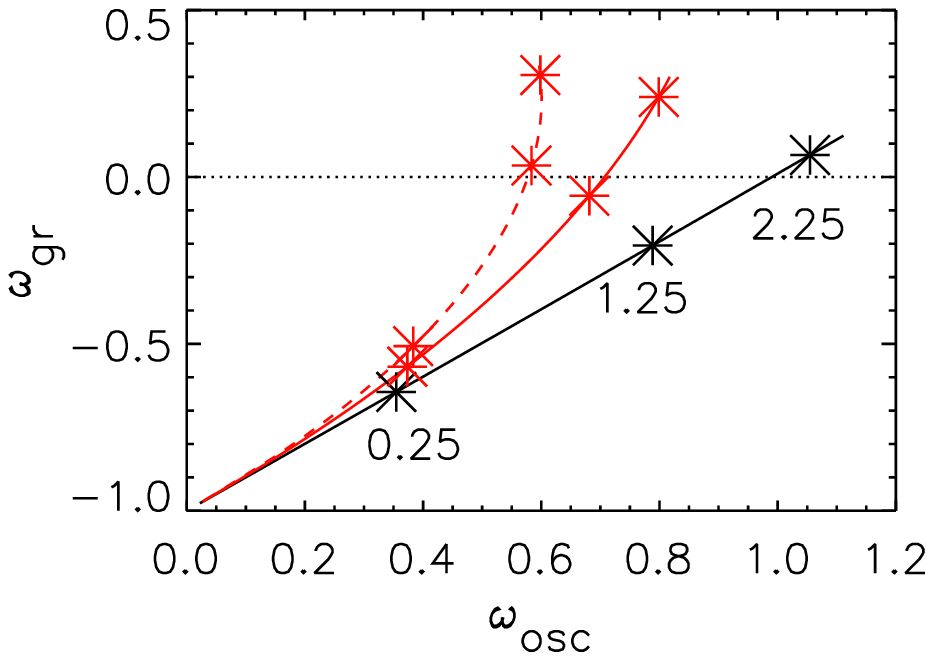}
\includegraphics[width=1.1\columnwidth]{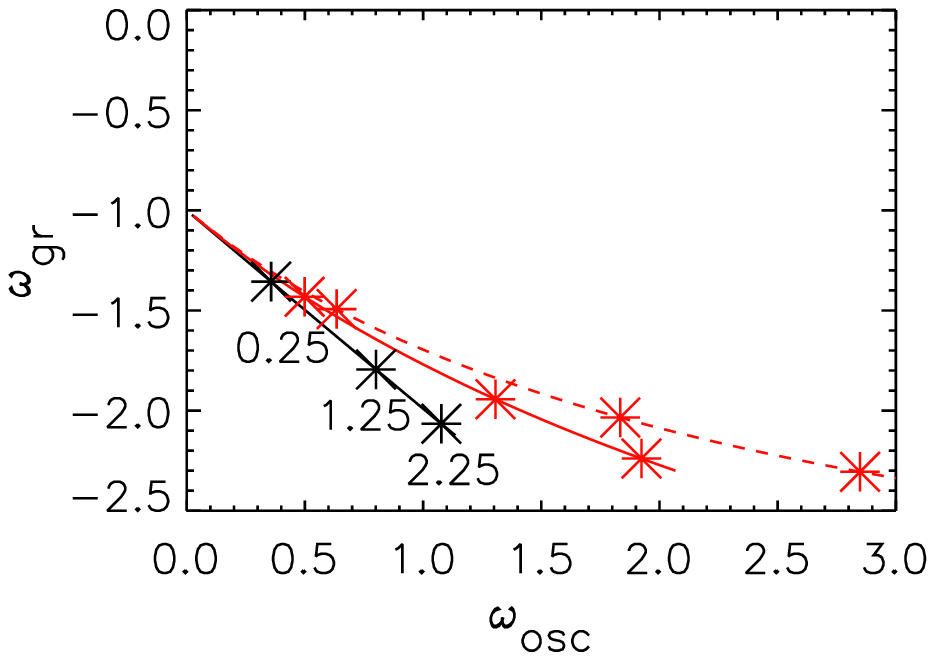}}}
\caption{The two solutions of the linear  Eq. (\ref{DD4}). Only the left solution provides unstable modes with $\omega_{\rm gr}>0$. Black line: $\tauc=0$, red solid: $\tauc=0.5$, red dashed: $\tauc=1$. The curves are marked with the values of $D$. The eigenvalues of the linear equation along the dotted zero line for $\omega_{\rm gr}$ are $D=2$ (black line), $D=1.33$ (red line) and $D=1$ (red dashed)}.
\label{fig11}
\end{figure*}

We start with $\Phi=1$  so that from  (\ref{DD3})
\begin{eqnarray}
\omega^2 + (2  +\i \tauc D)\omega + 1- \i D =0
\label{DD4}
\end{eqnarray}
 follows. Then we find a neutral solutions $\omega_{\rm gr}= 0$
for $D=2/(1+\tauc)$ and
\begin{eqnarray}
\omega_{\rm osc,1} = \frac{1}{1+\tauc}.  
\label{DD5}
\end{eqnarray}
The nonlocal $\alpha$ effect expression in Eq. (\ref{DD1}) reduces both the needed dynamo number for excitation as also the frequency of the oscillating $\alpha\Om$ dynamo.
Generally, the cycle frequency is not only determined by the diffusivity frequency but also on the value of the correlation time. For the rather long correlation time $\tauc=1$ the cycle time is twice the period for $\tauc=0$. Nevertheless, for any correlation time the cycle time exceeds the diffusion time of $2\pi$. We shall see that this prolonged cycle time  is the dominant effect of the  formulation (\ref{Ei4}) nonlocal in time. 

 If the correlation time fluctuates around a mean value then also the cycle time will fluctuate around a mean value - as also the magnetic cycle of the Sun does. After Eq. (\ref{DD5}) the normalized fluctuation of the cycle time corresponds to the normalized fluctuation of the correlation time.

%As expected, also the critical dynamo number is reduced by the finite correlation time. The line of zero growth rate $\omega_{\rm gr}$

The second frequency 
\begin{eqnarray}
\omega_{\rm osc,2} = - \frac{1+2 \tauc}{1+\tauc}\simeq -1 -\tauc
\label{D51}
  \end{eqnarray}
belongs to a  decaying  mode with $\omega_{\rm gr}<0$.

%%%%%%%%%%%%%%%%%%%%%%%
%%%%%%%%%%%%%%%%%%%%%%%
%\section{Nonlinear models}
%%%%%%%%%%%%%%%%%%%%%%%%%
We shall first  discuss the nonlinear model for a turbulence with a fixed correlation lengths  $\tauc$. 
%\begin{figure*}
%\centerline{
%\hbox{
%\includegraphics[width=0.7\columnwidth]{n001-AB.eps}
%\includegraphics[width=0.7\columnwidth]{n05a-AB.eps}
%\includegraphics[width=0.7\columnwidth]{n09a-AB.eps}}}
%\caption{The  solution of (\ref{D3}) with $\Psi$ after (\ref{D2}) for $D=2$ and for $\tauc=0.01$ (left),  $\tauc=0.5$ (middle)  and $\tauc=0.9$ (right).  The normalized field component $A$  (red) and the normalized field component $B$ (blue).  The cycle length grows with growing correlation time. For large $\tauc$ the variation of the maximal amplitudes remains small and also  the relative behaviours of $A$ and $B$ are almost independent of the correlation time. For $\tauc=0.5$ the cycle length $\approx 4.5$ diffusion times. The sign of $A$ depends on the sign of $C_\Om$, here $C_\Om=5$.}
%\label{fig3a}
%\end{figure*}

\begin{figure*}
\centerline{
\hbox{
\includegraphics[width=0.7\columnwidth]{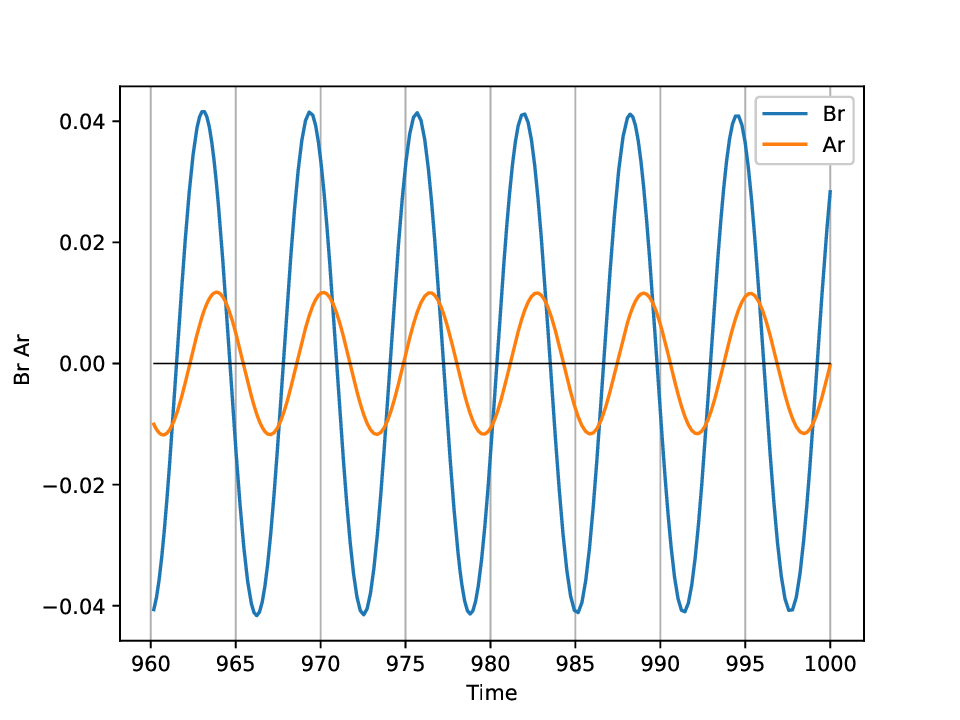}
\includegraphics[width=0.7\columnwidth]{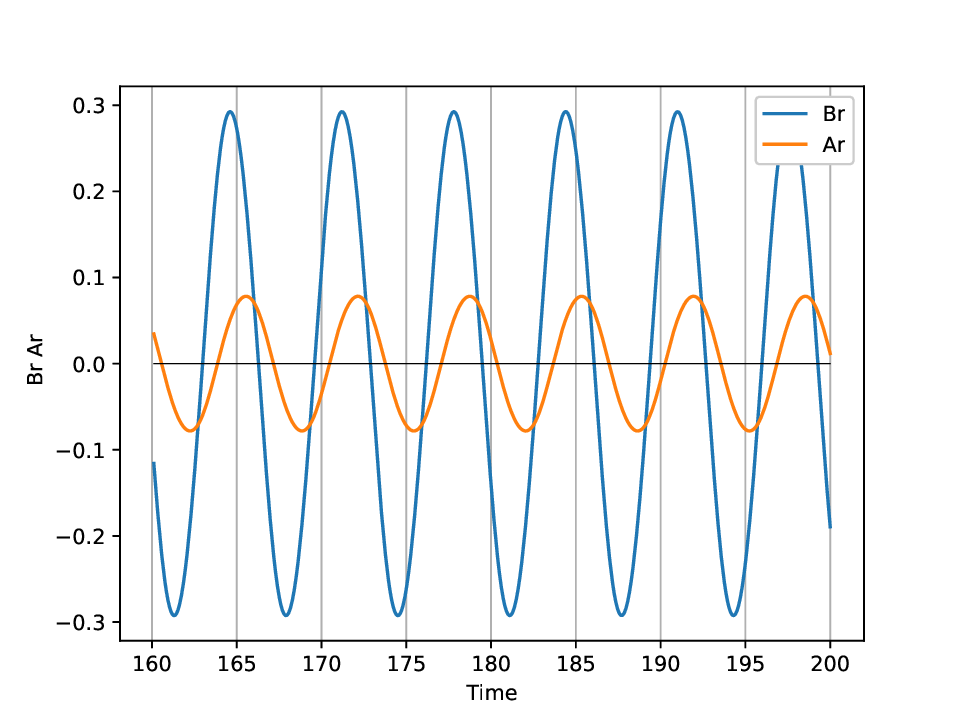}
\includegraphics[width=0.7\columnwidth]{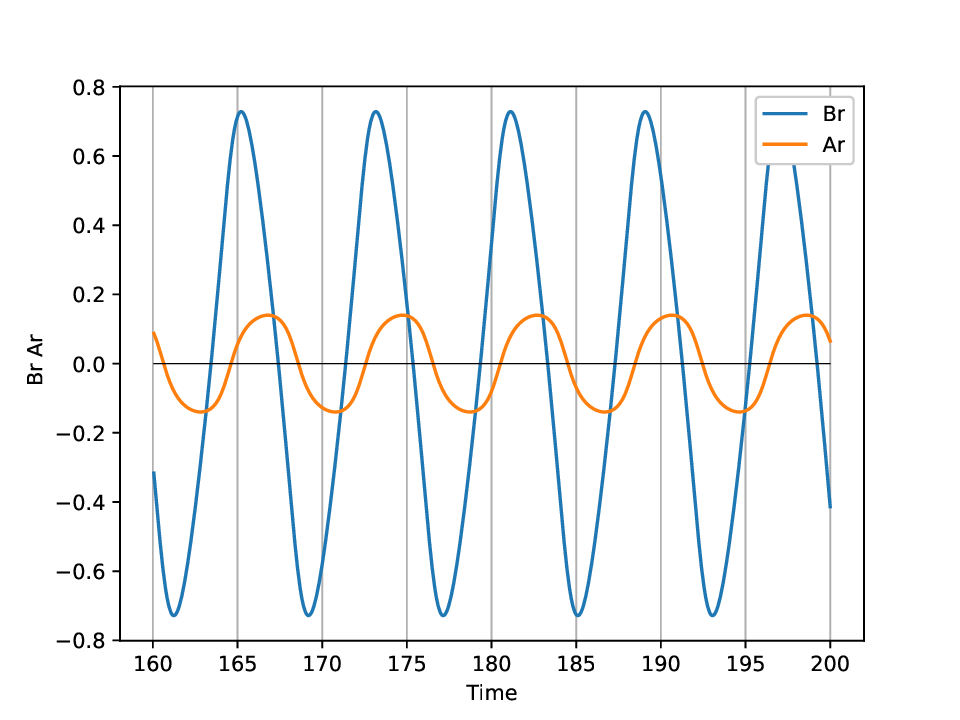}
}}
\caption{The  solution of Eqs. (\ref{D11}) for $D=2$. Left to right: $\tauc=0$, $\tauc=0.1$ and $\tauc=0.5$ .  The normalized field component $A$ (red) and the normalized (''azimuthal'') field component $B$ (blue).  The cycle length and the maximal amplitudes of $B$ grow with growing correlation time. The maximal amplitudes of $A$ remain almost independent on $\tauc$. Left to right: $\varepsilon=1, 0.85, 0.81$. $C_\Om=5$.}
\label{fig3b}
\end{figure*}

Positive $C_\Om$ indicates  uniform negative gradient of $\Om$. 
The quenching functions $\Phi$ and $\hat \Phi$ for $\tauc=0$ change into the 
 into the well-known result $1/(1+B^2)^2$ derived by \citep{R74} which is positive-definite. In contrast the common denominator in these expressions   for $\tauc>0$  may have a resonance somewhere along the line 
 \beg
 \frac{ \tauc{\dot B}}{B}= {2 },
 \label{reson}
 \ende
(with $|B|>1$, see Fig. \ref{fig00}). For small $\tauc$ this  condition describes a rather steep rise phase of the cycle. This phenomenon will have consequences for the resulting amplitudes of the dynamo solutions. Figure \ref{fig3b} gives  same models for the quasilinear parameter $D=2$ under application of the more complete quenching function. One only finds a weak dependence of the field amplitude on the correlation time  and the modest  increase of the cycle time with growing correlation times.

%%%%%%%%%%%%%%%%%%%%%

%%%%%%%%%%%%%%%%%%%%%%%%%%%%%%%
\subsection{Waldmeier profiles}
%We shall also discuss in detail the shape of the cycles of the toroidal fields.  
We now discuss in more detail the shape of the toroidal-field cycles.
While $\varepsilon=1$ corresponds to a symmetric half-cycle, values $\varepsilon<1$ describe cycles with faster rise than decline, as observed.
Figure \ref{fig3b} generally shows $\varepsilon<1$ for $\tauc>0$. 
This empirical asymmetry in the solar cycle is commonly associated with the Waldmeier rule \citep{W36}.
%The empirical realisation of this asymetry for the solar cycle has been called the Waldmeier rule \citep{W36}. 
%The profile of the half-cycles becomes the more  asymmetric the more important the $\dot B$-term in the formulation of  the $\alpha$effect is. 
The profile of the half-cycles generally becomes more asymmetric as the $\dot B$ term in the formulation of the $\alpha$ effect becomes more important.
This is understandable as by the first equation of (\ref{D11}) for nonvanishing correlation time the relation $\dot B \propto B$ follows $B$ as always positive(negative) in an upper(lower)  half-cycle so that the  rising branch must be steeper than the following declining branch. 
%This formal interpretation is confirmed by the run of the $\varepsilon$ with $\tauc$ in Fig. \ref{fig3b}. 
This interpretation is consistent with the behaviour of $\varepsilon$ as a function of $\tauc$ in Fig.~\ref{fig3b}.
%The sawtooth-profile (``Waldmeier profile'') of the cycles in  Fig. \ref{fig3b} is an immediately consequence of the nonlocal formulation of the $\alpha$ effect in Eq. (\ref{Ei4}).
The weak sawtooth profiles (``Waldmeier profiles'') of the cycles in Fig.~\ref{fig3b} arises naturally from the nonlocal formulation of the $\alpha$ effect in Eq.~(\ref{Ei4}), but the effect only appears for the longest correlation times.
As shown below, nonlinear effects alone can also generate this asymmetry in  supercritical models.
\begin{figure*}
\centerline{
\hbox{
\includegraphics[width=0.70\columnwidth]{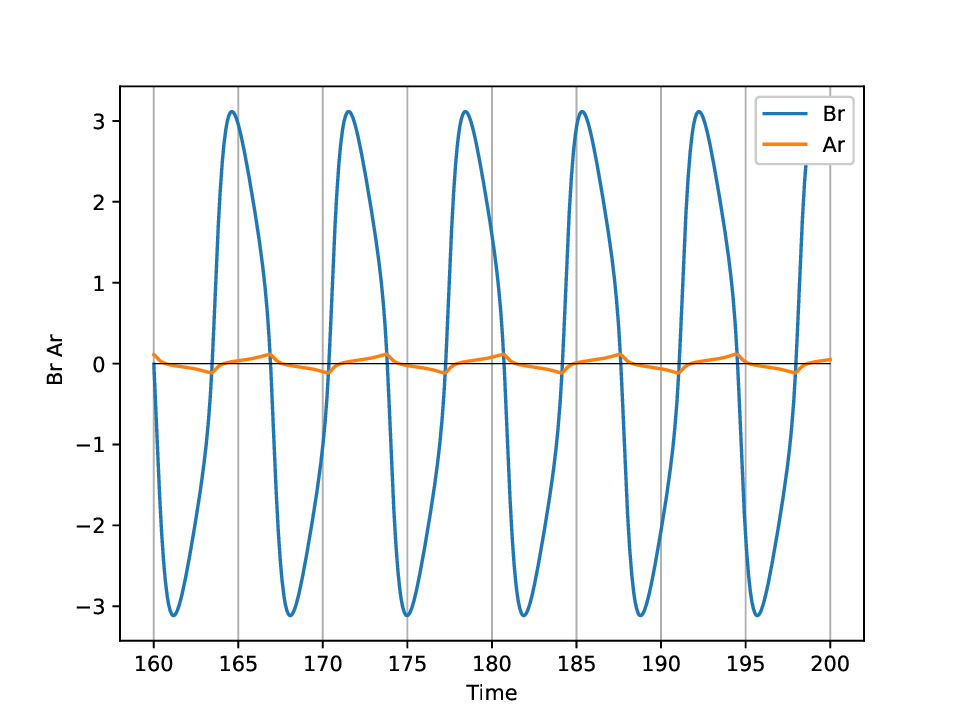}
\includegraphics[width=0.7\columnwidth]{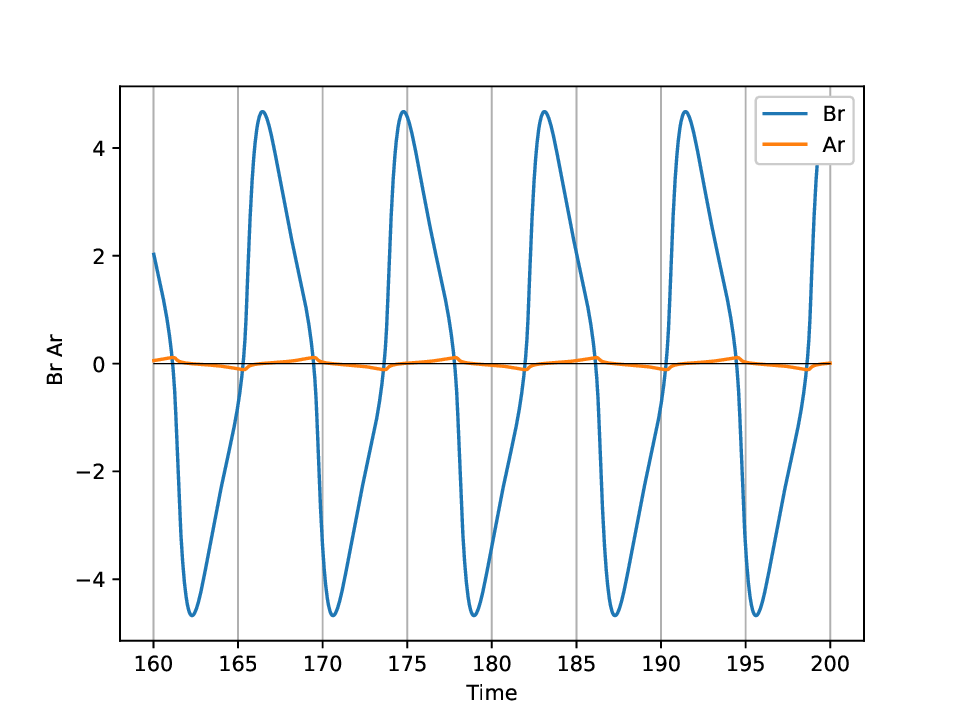}
\includegraphics[width=0.7\columnwidth]{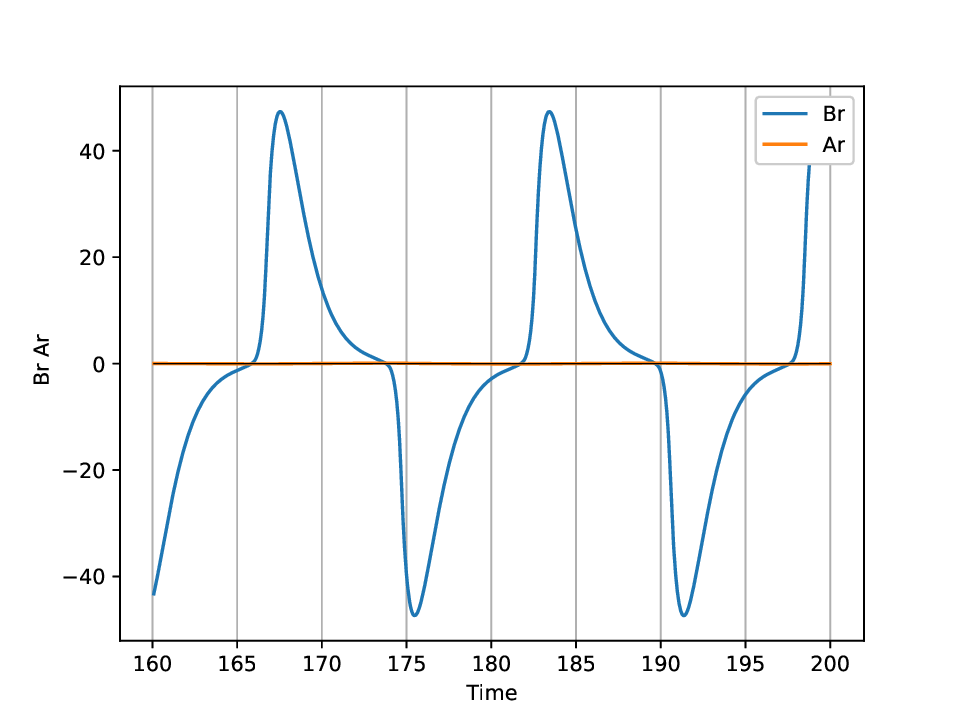}
}}
\caption{Similar to Fig. \ref{fig3b} but for the supercritical value $D=10$. The asymmetry of the rising and the declining parts of the cycles already exists for $\tauc=0$. The amplitudes and the cycle times  of $A$ and $B$ do not vary  from cycle to cycle. Left to right: $\varepsilon=0.57, 0.37, 0.34$. $C_\Om=25$.}
\label{fig3c}
\end{figure*}
\begin{figure*}
\centerline{
\hbox{
\includegraphics[width=0.7\columnwidth]{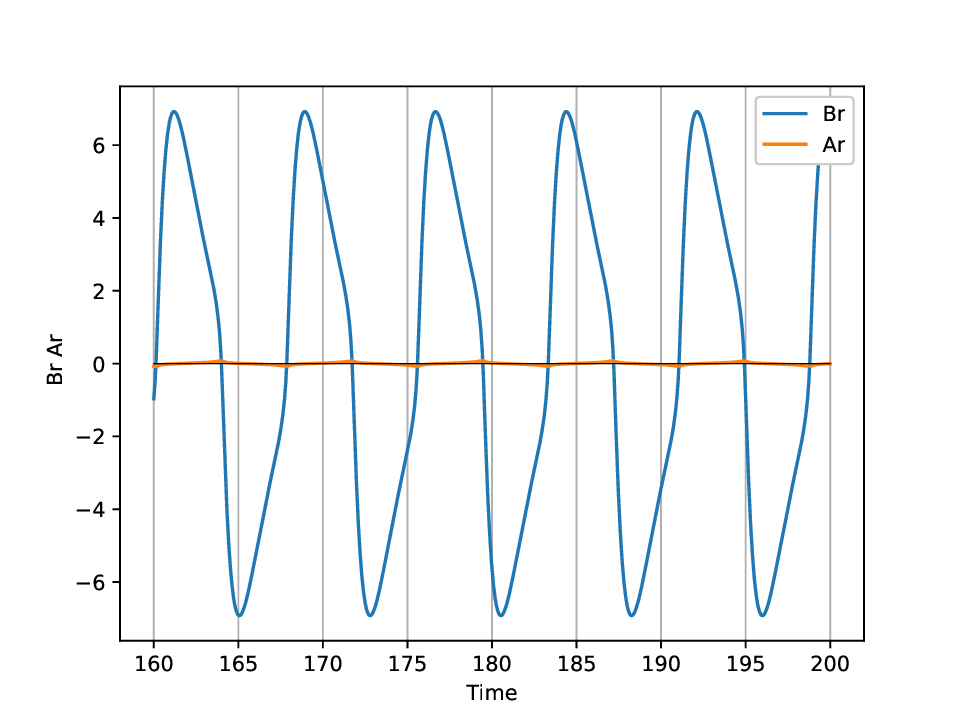}
\includegraphics[width=0.7\columnwidth]{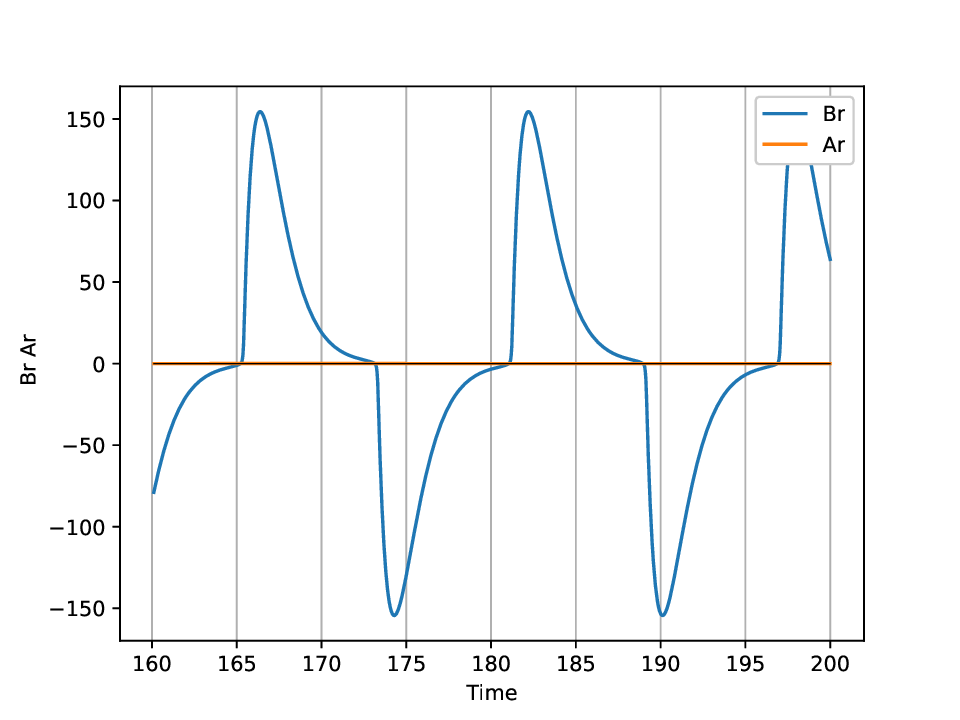}
\includegraphics[width=0.7\columnwidth]{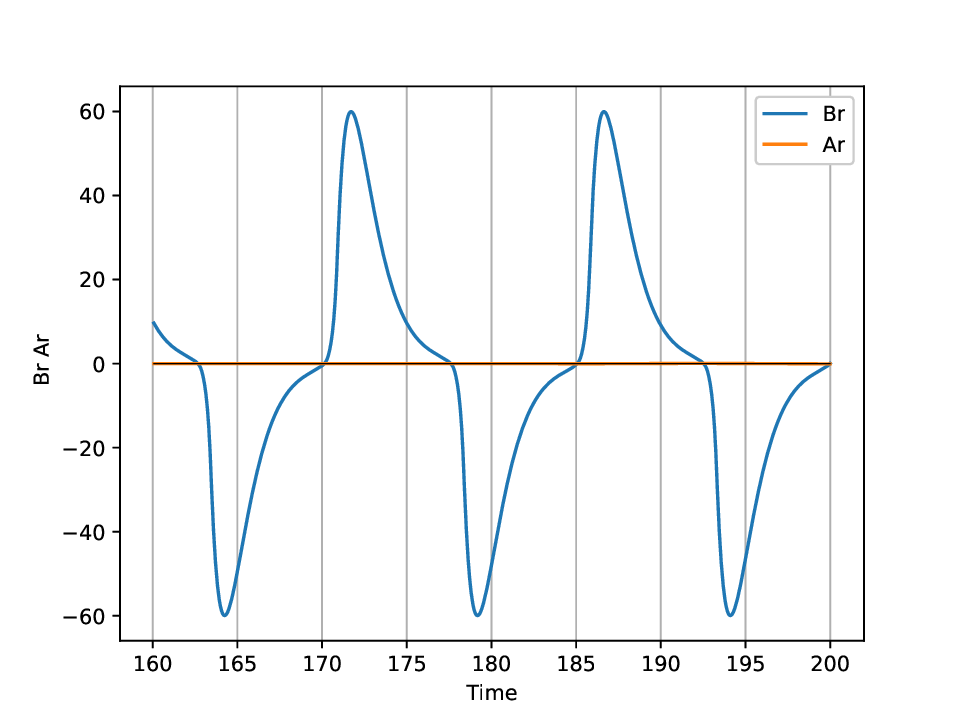}
}}
\caption{Similar to Fig. \ref{fig3b} but for the supercritical value $D=30$.  We note the strong increase of the magnetic amplitude of $B$ for $\tauc=0.1$. Left to right: $\varepsilon=0.34, 0.18,0.27$. $C_\Om=75$.}
\label{fig3d}
\end{figure*}

It remains to study the behaviour of the model for supercritical excitation. To this end the $\alpha$ effect is fixed to 
$C_\alpha=0.40$ while various values of $C_\Om$ are chosen mimicing growing rotation rates. The lower one, $C_\Om=5$, leads to a  value of $D=2$  close to the critical value for the linear dynamo (Fig. \ref{fig3b}). 

The second one, $C_\Om=25$, leads to a  supercritical excitation with $D=10$ where the interaction of several temporal modes should provide  asymmetric cycle shapes (Fig. \ref{fig3c}). 
Figure \ref{fig3d} provides the solutions for the highly supercritical value $D=30$. The questions are, how strong possible asymmetries are due to the nonlinearity and/or due to the new nonlocal term of the $\alpha$ effect or whether they may cancel each other.

A comparison of the left panels of these plots shows that a Waldmeier-type profile of the magnetic cycle can already arise in the local formulation of the $\alpha$ effect when the dynamo is  supercritical.
%Surprisingly, a comparison of the  left plots in Figs. \ref{fig3c} and\ref{fig3d} demonstrates the generation of an Waldmeier profile  of the magnetic cycle for the local formulation of the $\alpha$ effect but for  supercritical excitation. 
%The rise time becomes  shorter than the decline time even for $\tauc\to 0$, or, with other words, the maximum/minimum is  located closer to the reversal (zero) time  than in the middle of the reversals. It is a strong effect with  $\varepsilon \lsim 0.5$.  
This is already a strong effect, with $\varepsilon \lsim 0.57$.
The rise time becomes shorter than the decline time even for $\tauc \to 0$; in other words, the maximum or minimum is located closer to the reversal time than to the midpoint between successive reversals.
It does not exist for the linear or quasilinear solution of the dynamo equation as shown with the first plot of Fig. \ref{fig3b}. 
%The formal reason for the asymmetry is that after (\ref{D3}) the second derivative of $B$ depends on $\dot B$  - with the same sign between maximum and minimum - and $B$ with different signs between maximum and minimum. 

Formally, the asymmetry arises because in Eq.~(\ref{DD3}) the second derivative of $B$ depends on both $\dot B$, which keeps the same sign between a maximum and the subsequent minimum, and on $B$, whose sign changes over the same intervall.
%If our simplified $\alpha\Om$ dynamo model provides typical dynamo solutions then the decline  time in any cycle is never shorter  than the rise  time otherwise the cycle is not generated by an oscillating dynamo.
If the behaviour of our simplified $\alpha\Omega$ model is representative, then oscillatory dynamo solutions should generically exhibit decline times that are not shorter than the corresponding rise times.
%As also shown by Fig. \ref{fig3b} for the critical dynamo system the part of the $\alpha$ effect  for finite correlation time also produces cycles with shorter rising times  than decline times. 
As shown by Fig.~\ref{fig3b}, finite correlation time also produces cycles with shorter rise times than decline times in the near-critical dynamo regime. In this case, however, the effect remains modest.
%The effect, however, is small.  
For the (rather long) correlation time $\tauc=0.5$  the maximum already occurs after 45\% of the total time of the half-cycle while the decline phase  lasts 55\%.

%As shown by Fig. \ref{fig3c} another situation holds for supercritical excitation. 

Figure~\ref{fig3d} shows that the situation changes for highly supercritical excitation.
Here the asymmetry of the cycles is strong already for shorter correlation time. For $\tauc=0.1$ the rise time is only 15\% of the total time of the half-cycle. 
In this regime, nonlinearity and nonlocality of the $\alpha$ effect act together to enhance the Waldmeier-type asymmetry of the activity cycles.
%Nonlinearity and nonlocality of the $\alpha$effect are supporting each other to produce the asymmetry of the Waldmeier-type in the temporal form of the activity cycles. 
As can be seen from the series plots in Fig.  \ref{fig3d}, this cooperative effect weakens again for very large values of the correlation time. A similar trend is found for the magnetic-field amplitude $B$, whereas the cycle time continues to increase.
%As one can take from the series plots in Figs. \ref{fig3b} and  \ref{fig3c} this supporting action ends for very large values of the correlation time. The same holds for the magnetic field amplitude $B$ while the growth of the cycle times continues. 

%Figure \ref{fig3d} provides a new phenomenon concerning the field amplitudes peaking at $\tauc\simeq 0.1$. 
Figures~\ref{fig3d} shows another noteworthy feature, namely a distinct maximum of the field amplitude near $\tauc \simeq 0.1$.
For both shorter or longer correlation times the generated fields are much weaker than for  $\tauc\simeq 0.1$. 
%We shall discuss the reason and the consequences of this numerical finding in the following.
In the following we shall discuss the origin and possible consequences of this numerical result.

%%%%%%%%%%%%%%%%%%%%%%%%%%%%%%
\subsection{Influence of the dynamo number $D$}
\begin{figure}
\centerline{
%\hbox{
\includegraphics[width=1.0\columnwidth]{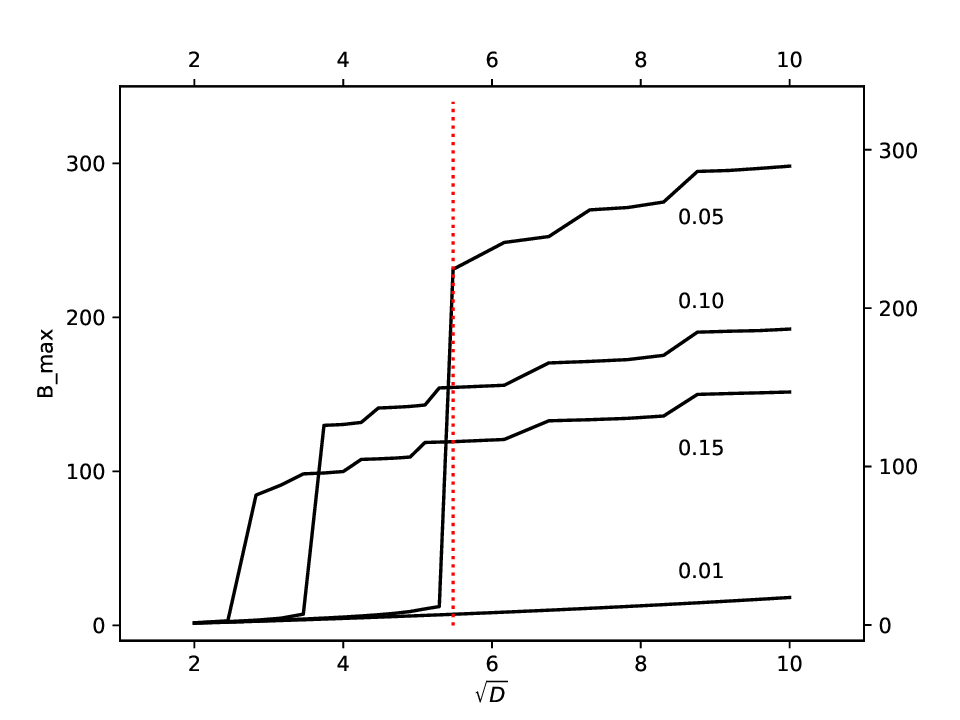}
}
\caption{The peak amplitudes of the magnetic field $B$  for various values of $\tauc$ as function of    $\sqrt{D}$. As for other models with different quenching functions the exponent $m$  is $m=0.5$. The lines are marked with their value of $\tauc$. One finds $D_{0}\simeq 1.4$. The vertical line belongs to $D=30$  which defines the models  in Fig. \ref{fig3d}. }
\label{fig9}
\end{figure}

%%%%%%%%%%%%%%%%%%%%%%%%%%%%%%%%%%%%%%%%%%%%%%%%%%%%%%%%%%%%
%It remains to study the influence of the dynamo number $D$ on the results of the nonlinear equation (\ref{D3}). 
It remains to study the influence of the dynamo number $D$ on the solutions of the nonlinear equation (\ref{DD1}).
%This is a nontrivial problem as both the time-derivatives of $B$ and  the magnetic field itself quench the system. 
This is a non-trivial problem, since both the time derivative of $B$ and the magnetic field itself contribute to the quenching of the system.
%The question is whether the fields  follow a relation
The question is whether the field amplitudes follow a relation of the form
\begin{eqnarray}
B_{\rm max} \simeq \beta\   (D-D_{0})^m,
\label{D7}
\end{eqnarray}
%(with $D_{0}$ the critical value of $D$ where the dynamo starts) with  $m\simeq 0.5$ as known from models with the standard quadratic $\alpha$ quenching \citep{SS1989}. 
where $D_{0}$ is the critical value of $D$ for dynamo onset, and $m\simeq 0.5$ is the familiar scaling known from models with standard quadratic $\alpha$ quenching \citep{SS1989}.
The Fig. \ref{fig9} demonstrates that the same result indeed follows for our much simpler  model with the new quenching functions (\ref{D12}) and (\ref{D13}) 
%but only for high enough dynamo numbers. 
%Equation (\ref{D7}) only holds for dynamo numbers exceeding a critical one, $D_{\rm crit}$, which increases for decreasing $\tauc$. On the other hand, the induced value of $B_{\rm max}$ at $D_{\rm crit}$ grows with $D_{\rm crit}$. 
%The  curves in Fig. \ref{fig9} show a particular behaviour. 
The curves in Fig.~\ref{fig9} summarize the results of a very high number of models. They show a characteristic behaviour. The magnetic field amplitudes of the cycles for a fixex correlation time $\tauc$ grow with $\sqrt{D}$ but there is a weak-field branch and a strong-field branch seperated by a steep jump at a certain dynamo number which is larger for shorter correlation times (at infinity for $\tauc=0$). The same holds for the level of the amplitudes of the strong-field branch which is highest for the small $\tauc$.
%For high enough dynamo numbers $D>36$ the magnetic amplitudes are much higher for $\tauc\simeq 0.05$ than for smaller or higher values. 

We note as an example that for sufficiently large dynamo numbers, $D>28$, the magnetic amplitudes are significantly larger near $\tauc \simeq 0.05$ than for either smaller or larger correlation times. Hence, for fixed dynamo number $D$, the induced amplitudes $B_{\rm max}$ of the toroidal field will depend quite sensitively on the correlation time $\tauc$.
%There is thus a strong dependence for fixed dynamo number $D$ of the induced magnetic amplitudes $B_{\rm max}$ on the correlation time $\tauc$. 
%Any deviation of $\tauc$ from the value 0.05 will lead to cycles of {\em weaker} amplitudes.
Any departure of $\tauc$ from the value $0.05$ leads to weaker cycle amplitudes in this model. Accordingly, even modest fluctuations of the correlation time may produce substantial variations in the amplitude of the $B$-field cycle.
 %Any small fluctuatio of the correlation time will lead to large fluctuations of the amplitudes of the $B$-field cycle.  
% Obviously, the  supercritical system (\ref{D11}) possesses a weak-field domain and a strong-field domain for a narrow domain of correlation times defined by  (\ref{reson}). 
In this sense, the supercritical system (\ref{DD1}) exhibits a weak-field domain and a strong-field domain over a relatively narrow interval of correlation times associated with condition (\ref{reson}).
The form of (\ref{D7}) with the square-root-dependence, however,   does not depend on the structure of the quenching functions. We also note, that the  magnetic field component $A$ does neither depend on the dynamo number $D$ nor on the correlation time.

 \subsection{Correlation time variations}
%%%%%%%%%%%%%%%%%%%%%%%
%Another effect is found for nonstationary correlation times. 
A further effect appears when the correlation time is allowed to vary in time.
%For simplification only  a stepwise approximation for its time-dependence 
%For simplification only  a stepwise approximation for its time-dependence 
%s used instead of a random function.
For simplicity, we use a stepwise approximation for its time dependence rather than a fully random function.
%\begin{eqnarray}
%\tauc=\tau_0  + \tau_1 \sin\omega_\tau t
%\label{D6}
%\end{eqnarray}
%is used instead of a random function. 
%Here $\tau_1$ gives the amplitude and $\omega_\tau$ gives the frequency of the correlation time variation.
%In order to study the influence of a fluctuating correlation time we probe the frequency $\omega_\tau$ in Eq. (\ref{D6}) with low values $\omega_\tau=0.05$, i.e.  long-term variations with a period of $40\pi$. 
%to large values $\omega_\tau=0.2$ (short-term variations). 
%The total variation $\delta\tauc$ is of the order of $\tauc$, i.e. $\tau_0=0.1$ and  $\tau_1=0.09$. 
% High-frequency variations of the correlation time are fully damped out.There are regular but rather weak variations of the cycle times. Positive cycles and negative cycles are not  symmetric to the zero-line. 
 High-frequency variations of the correlation time are strongly damped. The resulting variations of the cycle times are regular but weak. Positive and negative cycles are no longer exactly symmetric with respect to the zero line.
\begin{figure}
\centerline{
\vbox{
\includegraphics[width=1.0\columnwidth]{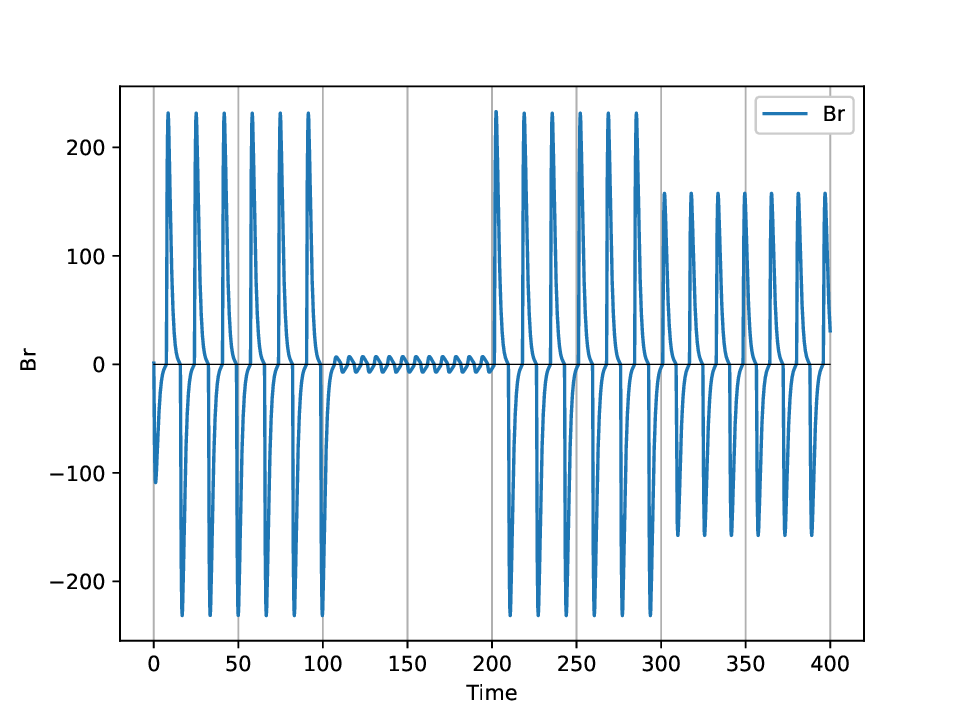}
\includegraphics[width=1.0\columnwidth]{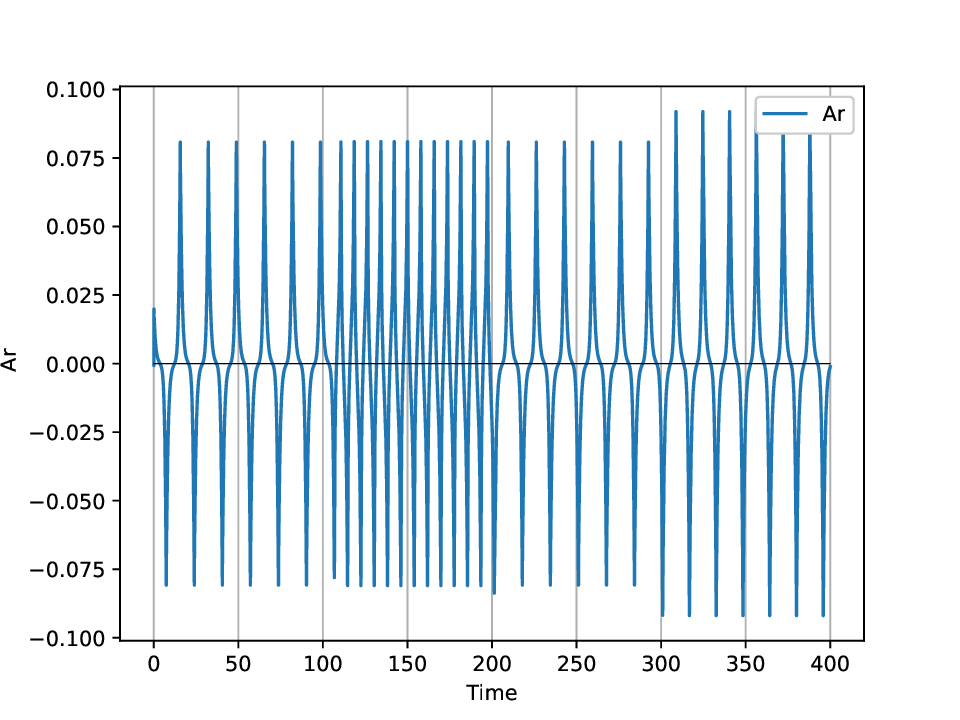}
}}
\caption{The dynamo solution for $B$ (top) and $A$ (bottom) for stepwise correlation  time $\tauc$ fluctuating around $\tau_0=0.05$ as described in the text. 
 $D=30, C_\Om=75 $.}
\label{fig4b}
\end{figure}
Here we concentrate to variations around the correlation time of $\tauc=0.05$. For  supercritical dynamo numbers $D=30$ the dynamo-generated fields $B$ are weaker for $\tauc\neq 0.05$ than for $\tauc=0.05$. The fluctuations are defined  by $|\delta \tauc|=0.04$. For 100 timesteps the fluctuation may vanish,  $\delta \tauc=0$, then for 100 timesteps $\delta \tauc=-0.04$ is required followed by again 100 timesteps  $\delta \tauc=0$ and then 50 timesteps $\delta \tauc=0.04$. As it should be the sum of the fluctuation vanishes.
%The results for the short-term variation with  $\omega_\tau=0.2$ are shown by Fig. \ref{fig4a}. One finds high-amplitude cycles alternating with longer periods  and low-amplitude cycles with shorter periods. The are two  frequencies differing from the basic one: a variation due to the the variation of the correlation time with a period of $10\pi$ and a rather long variation with a period of about 500 diffusion times.
Figure \ref{fig4b} gives the solution of Eq. (\ref{DD1}) for correlation times fluctuating as described between $\tauc=0.01$ and $\tauc=0.09$. 
%Both cycle times and amplitudes vary with time. 
Both cycle times and amplitudes vary in time. 
%The maximum values of $B$ and the cycle period are weakly correlated, in the sense that larger amplitudes tend to be associated with somewhat longer cycles.
The maximum values of $B$ and the cycle period are weakly correlated, in the sense that larger amplitudes tend to be associated with somewhat longer cycles.
%The maximum values of $B$ and the cycle period are  slightly correlated, i.e. the larger the amplitude the longer  the cycle. The magnetic cyles are  symmetric to $B=0$. 
%The lowest amplitudes and the middle amplitudes are due to the variation of the correlation time. We note that the two minima do not have the same depths.
%The low-amplitude episodes arise from the imposed variation of the correlation time. 
%We note that the two minima do not reach the same depth.

The low-amplitude episodes in this experiment arise from the imposed variation of the correlation time.We also note that the two minima do not reach the same depth, only the amplitudes for negative $\tauc$ are rather deep. We note that the variation  of the maximal amplitudes $A$ is only very small, grand minima only happen for the toroidal field $B$.
Obviously,  modest temporal variations of the turbulent correlation time may provide a natural mechanism for grand minima modulations  in nonlinear dynamo models.

%%%%%%%%%%%%%%%%%%%%%%%%%%%%%%%%%%%%%%%%%%
%\section{Results and discussion}
%
%%%%%%%%%%%%%%%%%%%%%%%%%%%%%
%\begin{figure*}
%\centerline{
%\hbox{
%\includegraphics[width=0.7\columnwidth]{n01-Br2-Ar2.eps}
%\includegraphics[width=0.7\columnwidth]{p01-Br2-Ar2.eps}
%\includegraphics[width=0.7\columnwidth]{p51-Br2-Ar2.eps}}}
%\caption{The magnetic energies $B^2$ of the solutions for  short correlation time $\tauc=0.1$ with growing dynamo numbers $D=2$ (left), $D=10$ (middle) and $D=30$ (right)}
%\label{fig5}
%\end{figure*}

%The upper panel of Fig. \ref{fig4b} diplays cycles of the toroidal field $B$ with  rise times and decline times which are  plotted in Fig. \ref{fig4c}. 

One finds the rise time almost constant in time while the decline times are reduced in the minima. Hence, the cycle time as the sum of both numbers is shorter in the minima rather than in  epochs with unchanged correlation time. Cycle time and cycle amplitude are thus slighty correlated. 
In these solutions, the decline time is more sensitive to the correlation time normalized with the magnetic-diffusion time, whereas the rise time depends much less strongly on the actual value of the correlation time.
%We note that only the decline time reflects the length of the correlation time normalized with the magnetic-diffusion time while the rise time does not depend on the value of the actual correlation time. 
%The asymmetry factor $\varepsilon$ with about 0.25 is very low indicating a clear sawtooth profile of the cycle. In contrast to the maxima of $A$ those for the latitudinal field $B$ are much stronger. For the ratio $r= B/A$ the values vary between $r\simeq 160$ and $r\simeq 250$ certainly depending on the dynamo number of the model. 
The asymmetry factor $\varepsilon$, with values around 0.25, is very low and indicates a clear sawtooth profile of the cycle.
In contrast to the maxima of $A$ those for the latitudinal field $B$ are much stronger. For the ratio $B/A$ the values vary between $\simeq 100$ and $\simeq 1000$ certainly depending on the dynamo number of the model.

%The influence of the correlation time fluctuations on the latitudinal field $A$ is much weaker than on the longitudinal field $B$. 
The influence of the correlation-time fluctuations on the latitudinal field $A$ is much weaker than on the longitudinal field $B$.
%While the minima can clearly be observed with the field $B$, they only weakly appear in the runs of the $A$-field.
While the minima are clearly visible in the field $B$, they appear only weakly in the time series of the $A$ field.

%%%%%%%%%%%%%%%%%%%%%%%%
\section{Conclusions}
%%%%%%%%%%%%%%%%%%%%%%%%%
In this paper we have explored the consequences of extending the traditional $\alpha$ effect by a term nonlocal in time, such that the turbulent electromotive force depends not only on the mean magnetic field but also on its temporal derivative. Within the framework of SOCA this leads, to leading order, to an expression of the form
\begin{eqnarray}
	{\vec{\cal E}} \propto \alpha \left( \bar{\vec B} - \tau_{\rm corr} \dot{{\vec B}} \right),
\end{eqnarray}
together with a derivative-dependent quenching term. In physical terms, the induction process then retains a finite memory of the recent magnetic-field evolution, and the standard local formulation is replaced by one in which the turbulence responds to a field that is effectively sampled in the past.

The first consequence of this modification is that the excitation conditions and the cycle periods of dynamo models are altered in a systematic way. In the linear problem, the correlation time acts as an additional control parameter: the threshold for dynamo excitation is reduced relative to the local case, while the oscillation period is increased. The nonlocal term therefore changes not only the quantitative values of the eigenfrequencies, but also the qualitative structure of the solution space, in particular through the appearance of distinct regimes separated by the condition $\tau_{\rm corr} C_\alpha \simeq 1$.

In the nonlinear regime, the most important result is that the temporal derivative entering the $\alpha$ effect naturally breaks the symmetry between rising and declining phases of the cycle. This leads to oscillatory solutions with shorter rise times than decline times and therefore to the familiar sawtooth-shaped cycle profiles. Near criticality this asymmetry follows directly from the nonlocal contribution. In more strongly supercritical models, however, nonlinear effects alone can already generate some degree of asymmetry even in the local formulation, and the nonlocal term then acts to enhance this behaviour further. The present results therefore suggest that temporal nonlocality is not necessarily the only source of cycle asymmetry, but it provides a simple and physically motivated mechanism that strengthens it and extends it over a broad parameter range.

A second important result is that the model reproduces Waldmeier-type trends in a natural way. As the correlation time increases, the cycle asymmetry becomes stronger and, in suitable parameter ranges, the rise rate and cycle amplitude become positively correlated. In the present formulation, both the field amplitude and the rise time may increase with increasing $\tau_{\rm corr}$, but the amplitude grows faster, so that the rise-rate--amplitude relation acquires the correct sense. By contrast, a negative correlation between rise time and cycle amplitude does not emerge from the present simplified model. This suggests that the finite correlation time of the turbulence may be sufficient to account for part of the observed phenomenology, but perhaps not for all of it.

The nonlinear solutions also show that, for sufficiently supercritical dynamo numbers, the magnetic-field amplitude can depend rather sensitively on the value of the correlation time. In particular, the field amplitude reaches a maximum at an intermediate value of $\tau_{\rm corr}$, while both smaller and larger values lead to weaker solutions. This implies that the turbulent correlation time does not act merely as a monotonic control parameter, but rather selects an optimal regime for field amplification. Such a dependence is particularly interesting because it opens the possibility that even modest temporal variations of $\tau_{\rm corr}$ may induce substantial long-term modulation of cycle amplitudes.

When the correlation time is allowed to vary in time, this possibility is indeed borne out by the model. Even in the simplified stepwise realizations studied here, the dynamo develops alternating epochs of stronger and weaker cycles, together with accompanying variations in the cycle period. The modulation remains much more pronounced in the toroidal than in the poloidal component, and the minima do not in general reach the same depth. Although the present experiments are still idealized, they suggest that temporal variability of the turbulent memory time may provide a plausible route toward weak-cycle episodes reminiscent of grand minima.

The physical interpretation of these results is that dynamo saturation is no longer governed by field amplitude alone, but also by the rate at which the large-scale field changes over the correlation time of the turbulence. In this sense, the nonlocal formulation provides a simple mechanism for producing finite-memory effects, cycle asymmetry, and long-term modulation within a low-order mean-field description. This conclusion is of interest not only for solar dynamos but also for stellar activity cycles more generally, because asymmetric rise and decay are observed in many late-type stars and may reflect the same generic finite-memory effect.

The model also points to possible observational consequences. If the effective source term depends on both the magnetic amplitude and its temporal derivative, then the efficiency of large-scale field regeneration should vary with cycle phase, with the strongest suppression expected during epochs of rapid field evolution. In the solar context, this may influence the skewness of activity cycles, the rise-rate--amplitude relation, and possibly the lag between activity maximum and polar-field reversal. In a Babcock--Leighton interpretation, one may further expect that source proxies based on active-region tilt or dipole moment correlate not only with activity level but also with its temporal derivative. In stellar activity data, the same mechanism may appear statistically through correlations between cycle asymmetry, amplitude, and period.

The present analysis is based on highly simplified dynamo models, and the results should therefore be interpreted primarily at a qualitative level. Nevertheless, the emergence of derivative-dependent saturation from a physically motivated nonlocal $\alpha$ effect is encouraging, especially in view of recent data-driven reconstructions of the solar cycle that also point to an explicit role of $\dot B$. A natural next step is therefore to incorporate the present nonlocal formulation into more realistic solar and stellar dynamo models, and to confront its predictions with long-term activity records and magnetic-field proxies.

\bibliographystyle{aa}
\bibliography{cycles}

\begin{thebibliography}{19}
\expandafter\ifx\csname natexlab\endcsname\relax\def\natexlab#1{#1}\fi

\bibitem[{{Bonanno} \& {Arlt}(2025)}]{2025SoPh..300..116B}
{Bonanno}, A. \& {Arlt}, R. 2025, \solphys, 300, 116

\bibitem[{{Cameron} \& {Sch{\"u}ssler}(2008)}]{CS08}
{Cameron}, R. \& {Sch{\"u}ssler}, M. 2008, \apj, 685, 1291

\bibitem[{{Charbonneau}(2020)}]{2020LRSP...17....4C}
{Charbonneau}, P. 2020, Living Rev. Sol. Phys., 17, 4

\bibitem[{{Garg} {et~al.}(2019){Garg}, {Karak}, {Egeland}, {Soon}, \&
  {Baliunas}}]{G19}
{Garg}, S., {Karak}, B.~B., {Egeland}, R., {Soon}, W., \& {Baliunas}, S. 2019,
  \apj, 886, 132

\bibitem[{{Hoyng} {et~al.}(2001){Hoyng}, {Ossendrijver}, \&
  {Schmitt}}]{Hoyng90}
{Hoyng}, P., {Ossendrijver}, M.~A.~J.~H., \& {Schmitt}, D. 2001, Geophysical
  and Astrophysical Fluid Dynamics, 94, 263

\bibitem[{{Lehtinen} {et~al.}(2016){Lehtinen}, {Jetsu}, {Hackman}, {Kajatkari},
  \& {Henry}}]{L16}
{Lehtinen}, J., {Jetsu}, L., {Hackman}, T., {Kajatkari}, P., \& {Henry}, G.~W.
  2016, \aap, 588, A38

\bibitem[{{Meinel} \& {Brandenburg}(1990)}]{Meinel90}
{Meinel}, R. \& {Brandenburg}, A. 1990, \aap, 238, 369

\bibitem[{{Moffatt}(1972)}]{M73}
{Moffatt}, H.~K. 1972, Journal of Fluid Mechanics, 53, 385

\bibitem[{{Ol{\'a}h} {et~al.}(2009){Ol{\'a}h}, {Koll{\'a}th}, {Granzer},
  {Strassmeier}, {Lanza}, {J{\"a}rvinen}, {Korhonen}, {Baliunas}, {Soon},
  {Messina}, \& {Cutispoto}}]{OKG09}
{Ol{\'a}h}, K., {Koll{\'a}th}, Z., {Granzer}, T., {et~al.} 2009, \aap, 501, 703

\bibitem[{{Parker}(1955)}]{1955ApJ...122..293P}
{Parker}, E.~N. 1955, \apj, 122, 293

\bibitem[{{Reinhold} {et~al.}(2017){Reinhold}, {Cameron}, \& {Gizon}}]{RCG17}
{Reinhold}, T., {Cameron}, R.~H., \& {Gizon}, L. 2017, \aap, 603, A52

\bibitem[{{R\"udiger}(1974)}]{R74}
{R\"udiger}, G. 1974, Astronomische Nachrichten, 295, 275

\bibitem[{{R{\"u}diger} \& {Arlt}(2003)}]{RA03}
{R{\"u}diger}, G. \& {Arlt}, R. 2003, in Advances in Nonlinear Dynamics, ed.
  A.~{Ferriz-Mas} \& M.~{N{\'u}{\~n}ez}, 147

\bibitem[{{R{\"u}diger} {et~al.}(2013){R{\"u}diger}, {Kitchatinov}, \&
  {Hollerbach}}]{RK13}
{R{\"u}diger}, G., {Kitchatinov}, L.~L., \& {Hollerbach}, R. 2013, {Magnetic
  Processes in Astrophysics: theory,simulations, experiments} ({Wiley-VCH})

\bibitem[{{Schmitt} \& {Sch\"ussler}(1989)}]{SS1989}
{Schmitt}, D. \& {Sch\"ussler}, M. 1989, \aap, 223, 343

\bibitem[{{Steenbeck} \& {Krause}(1966)}]{SK66}
{Steenbeck}, M. \& {Krause}, F. 1966, Zeitschrift Naturforschung Teil A, 21,
  1285

\bibitem[{Stix(2012)}]{stix2012sun}
Stix, M. 2012, The Sun: An Introduction, Astronomy and Astrophysics Library
  (Springer Berlin Heidelberg)

\bibitem[{{Waldmeier}(1935)}]{W36}
{Waldmeier}, M. 1935, Astronomische Mitteilungen der Eidgen{\"o}ssischen
  Sternwarte Zurich, 14, 105

\bibitem[{{Willamo} {et~al.}(2020){Willamo}, {Hackman}, {Lehtinen},
  {K{\"a}pyl{\"a}}, {Olspert}, {Viviani}, \& {Warnecke}}]{WHL20}
{Willamo}, T., {Hackman}, T., {Lehtinen}, J.~J., {et~al.} 2020, \aap, 638, A69

\end{thebibliography}

\begin{appendix}
\section{}
The MHD equations are attacked by the linear-in-time Fourier expressions (\ref{uibi})
%$\vec{\hat u}+ \vec{\tilde u} t$ and  $\vec{\hat b}+ \vec{\tilde b} t$ 
leading to the two sets of equations 
\beg
\begin{aligned}
&(-{\rm i} \omega
 + \nu k^2)\hat u_i +\tilde u_i=(-{\rm i} \omega
 + \nu k^2) \hat u_i^{(0)}+ B^{0}\hat b_i\\
& (-{\rm i} \omega
 + \nu k^2) \tilde u_{i}=B^{1}\hat b_i + B^{0}\tilde b_i 
 \end {aligned}
\label{A0}
\ende
and
\beg
\begin{aligned}
&(-{\rm i} \omega
 + \eta k^2)\hat b_i +\tilde b_i = \mu_0\rho B^{0}\hat u_i,\\
 &(-{\rm i} \omega
 + \eta k^2)\tilde b_i = \mu_0\rho( B^{1}\hat u_i+ B^{0}\tilde u_i).
 \end {aligned}
\label{A1}
\ende
The determinant of the equation system can be written as
\begin{eqnarray}
det = 
&\left((-{\rm i} \omega + \eta k^2)^2 - \frac{(\vec{k}\vec{B_0})^2}{\mu_0\rho}\right)^2+ \nonumber\\
&+\frac{1}{\mu_0\rho}\left(2 (-{\rm i} \omega+ \eta k^2) (\vec{k}\vec{B_0})- (\vec{k}\vec{\dot B})\right)^2
\label{A2}
\end{eqnarray}
($\Pm=1$), forming the sum of two squares. Wave solutions are thus excluded. It follows
\beg
\begin{aligned}
&\hat u_i = \frac{(-{\rm i} \omega + \eta k^2)^4}{det} \hat u^{(0)}_i,\\
&\hat b_i= {\rm i} \frac{(-{\rm i} \omega + \eta k^2)^2}{det}\left( (-{\rm i} \omega + \eta k^2) (\vec{k B_0}) - (\vec{k}\vec{\dot B}) \right) \hat u^{(0)}_i.
\end{aligned}
\label{A3}
\ende
Hence,
\beg
 \hat u_i \hat b^*_j =-\i\ (-\i \omega+\eta k^2)^4(\i \omega+\eta k^2)^2
 \frac{(\i \omega+\eta k^2) (\vec{ k B_0})- (\vec { k \dot B)}}{det \  det^*} \hat u_i^{(0)} \hat u_j^{(0)*}.\nonumber
 \label{A4}
 \ende
 Application of this result to an isotropic but helical turbulence as in \cite{R74} with the simplified spectrum $\delta(\omega)$ leads to the final result
 \beg
{\vec{\cal  E}}= C_\alpha \Psi ((1+B^2)^2(\vec{B}-\tauc \vec{\dot B}) + \tauc^2{\dot B}^2 \vec{B})
 \label{A5}
 \ende
 with 
  \beg
 \Psi = ((1+B^2)^2 +\tauc^2 {\dot B}^2 -4 \tauc B \dot B)^{-2}
 \label{A6}
 \ende
 for the turbulence-induced electromotic force. We note that the function $\Psi$ vanishes with (\ref{reson}). The quenching functions (\ref{D12}) and (\ref{D13}) are the immediate consequence.

\end{appendix}
\end{document}